# Anomalously large spin-dependent electron correlation in nearly half-metallic ferromagnet $CoS_2$

Hirokazu Fujiwara[1,2]*, Kensei Terashima[3], Junya Otsuki[4], Nayuta Takemori[4], Harald O. Jeschke[4], Takanori Wakita[4], Yuko Yano[1], Wataru Hosoda[1], Noriyuki Kataoka[1], Atsushi Teruya[5], Masashi Kakihana[5], Masato Hedo[6], Takao Nakama[6], Yoshichika Ōnuki[6,7], Koichiro Yaji[3], Ayumi Harasawa[2], Kenta Kuroda[2], Shik Shin[2], Koji Horiba[8], Hiroshi Kumigashira[8], Yuji Muraoka[1,4], and Takayoshi Yokoya[1,4]**

[1]*Graduate School of Natural Science and Technology, Okayama University, Okayama 700-8530, Japan*

[2]*Institute for Solid State Physics, The University of Tokyo, Kashiwa, Chiba 277-8581, Japan*

[3]*National Institute for Materials Science (NIMS), Tsukuba, Ibaraki, Japan*

[4]*Research Institute for Interdisciplinary Science, Okayama University, Okayama 700-8530, Japan*

[5]*Graduate School of Engineering and Science, University of the Ryukyus, Nishihara, Okinawa 903-0213, Japan*

[6]*Faculty of Science, University of the Ryukyus, Nishihara, Okinawa 903-0213, Japan*

[7]*RIKEN Center for Emergent Matter Science, Wako, Saitama 351-0198, Japan*

[8]*Photon Factory, Institute of Materials Structure Science, High Energy Accelerator Research Organization (KEK), 1-1 Oho, Tsukuba 305-0801, Japan*

*Correspondence to: fujiwara.h@s.okayama-u.ac.jp
**Correspondence to: yokoya@cc.okayama-u.ac.jp

**Abstract**

The spin-dependent band structure of $CoS_2$ which is a candidate for a half-metallic ferromagnet was investigated by both spin- and angle-resolved photoemission spectroscopy and theoretical calculations, in order to reappraise the half-metallicity and electronic correlations. We determined the three-dimensional Fermi surface and the spin-dependent band structure. As a result, we found that a part of the minority spin bands is on the occupied side in the vicinity of the Fermi level, providing spectroscopic evidence that $CoS_2$ is not but very close to a half-metal. Band calculations using density functional theory with generalized gradient approximation showed a good agreement with the observed majority spin $e_g$ bands, while it could not explain the observed band width of the minority-spin $e_g$ bands. On the other hand, theoretical calculations using dynamical mean field theory could better reproduce the strong mass renormalization in the minority-spin $e_g$ bands. All those results strongly suggest the presence of anomalously enhanced spin-dependent electron correlation effects on the electronic structure in the vicinity of the half-metallic state. We also report the temperature dependence of the electronic structure across the Curie temperature and discuss the mechanism of the thermal demagnetization. Our discovery of the anomalously large spin-dependent electronic correlations not only demonstrates a key factor in understanding the electronic structure of half-metals but also provides a motivation to improve theoretical calculations on spin-polarized strongly correlated systems.

# I. INTRODUCTION

Producing a source of spin-polarized carriers has been a challenging topic of research in solid state physics since the emergence of spintronics which manipulates the spin degree of freedom of carriers [1,2]. Efforts to realize carriers with large spin polarization have been vigorously undertaken in studies on half-metallic ferromagnets (HMFs) [3–7]. HMFs have metallic electronic structures with an energy gap at the Fermi level ($E_F$) for any one electronic spin state in the ground state, realizing 100% spin polarization at $E_F$.[8] Although many HMF candidates have been predicted for transition metal oxides and chalcogenides, Heusler alloys, and two dimensional materials,[9–13] $CrO_2$ has been the sole substance providing approximately 100% spin polarization in both point-contact Andreev reflection and spin-resolved photoemission spectroscopy experiments.[14–18] Further discovery and identification of half-metallic materials is necessary in order to understand characteristic phenomena of HMFs and to realize more thermally stable HMFs.

In $CrO_2$, there are two claims on electron correlation: a soft X-ray angle-resolved photoemission spectroscopy (ARPES) study shows that the observed majority spin band is very well explained by dynamical mean-field theory (DMFT) calculations assuming a small effective local Coulomb repulsion $U_{eff}$ < 1 eV.[19] On the other hand, in spin-resolved photoemission spectroscopy reports, the temperature dependence of the minority spin state near $E_F$ provides evidence that strong correlation effects are critical for understanding the electronic structure of $CrO_2$.[17,18] Although these claims seem to be contradictory, they can be interpreted consistently by assuming the existence of spin-dependent electron correlation. This spin-dependent electron correlation is expected to be a universal effect in ferromagnets, but few previous studies have investigated the effect.[20–22] In $CrO_2$, it is difficult to investigate both majority and minority spin bands simultaneously because there is no occupied minority spin band at $E_F$ at low temperatures.

$CoS_2$ is a ferromagnetic metal whose ordered moment is reported to be approximately 0.9$\mu_B$/Co with an easy-axis along the [111] direction.[23–25] The ordered moment is close to 1$\mu_B$/Co in a low-spin state of $Co^{2+}$ (configuration: $t_{2g}^6 e_g^1$), suggesting that $CoS_2$ has a half-metallic electronic structure including completely spin polarized $e_g$ bands crossing $E_F$. The half-metallicity has also been supported by theoretical band structure calculated within generalized gradient approximation (GGA), an experimental study of the de Haas–van Alphen (dHvA) effect, and an optical study measuring reflectivity of $CoS_2$.[10,24,26] On the other hand, the results of band calculations within local spin-density approximation (LSDA), point-contact Andreev reflection (PCAR), X-ray magnetic circular dichroism, and photoemission spectroscopy measurements indicated that $CoS_2$ has a nearly half-metallic electronic structure in which a bottom of the lowest-energy minority spin $e_g$ band touches $E_F$ or is slightly occupied near the R point.[10,27–32] This band-structure picture is also supported by the ARPES study on $CoS_2$[33] and PCAR studies on $Co_{1-x}Fe_xS_2$. In this system, the spin polarization at $E_F$ is enhanced by hole-doping.[34,35] In a polarized neutron-diffraction study, a non-half-metallic electronic state was inferred from the observed magnetization distribution, in which both $e_g$ and $t_{2g}$ states contributed to the magnetic moments.[36] This conflicts with the physical picture obtained by photoemission spectroscopy and density functional theory (DFT) studies based on LSDA.[31–33] So far, there is still no consensus on the half-metallicity of $CoS_2$. Therefore, a direct observation of spin-dependent electronic structure is required. If $CoS_2$ has an occupied minority spin band near $E_F$, we can observe both majority and minority spin bands simultaneously and discuss spin-dependent mass renormalization.

In order to reveal the spin-resolved electronic structure of $CoS_2$ experimentally, it has been pointed out that measurements with energy resolution less than 10 meV are needed. A photoemission study demonstrated that the peak width of a structure attributed to a minority spin state was 6.5 meV.[28] Clarifying the fine electronic structure with spin character can provide the decisive evidence for the half-metallicity of $CoS_2$.

In this study, we report the three-dimensional Fermi surface (FS) and the band structure of $CoS_2$ single crystals including the spin character determined by high-resolution spin- and angle-resolved photoemission spectroscopy (SARPES). We have found that both majority and minority spin bands cross $E_F$ along the ΓX line, unambiguously demonstrating that $CoS_2$ does not have a half-metallic electronic structure down to 20 K. Furthermore, the minority spin bands are renormalized by electron correlation effects, while the overall character of the majority spin band can be understood by the one-electron model. This indicates that spin-dependent electron correlation is important for understanding the electronic structure of $CoS_2$. Based on the experimentally determined electronic structure, we discuss (i) the half-metallicity in Sections II. A and B, (ii) the spin-dependent electron correlation effects from comparison with band structures computed in DFT based on GGA and in combination with DMFT (DFT+DMFT) in Sections II. C and D, (iii) the demagnetization process and the thermal decoherence of near-$E_F$ bands in Section II. E, and finally (iv) possible origins of the spin-dependent electron correlation in $CoS_2$ in Section II. F.

## II. RESULTS AND DISCUSSION
### A. Three-dimensional Fermi surface and spectral function

The photoemission spectroscopy measurements were performed for a cleaved surface of a CoS$_2$ single crystal. The low energy electron diffraction (LEED) pattern is shown in Fig. 1(a). Figures 1(c)–(f) show the experimental in-plane FSs for the cut P1 and P2 which are defined in Fig. 1(b). By determining the Fermi wave vectors ($k_F$s) and using two-fold symmetrization analysis (Figs. 1(e)-(h)), we found two large sheets and a pocket around the Γ point (α, β, and γ, respectively), one pocket around the X' point, and one pocket around the M point, in the cut P1.

Figures 2 (a) and (b) show the observed spectra along the ΓX line. As shown in Fig. 2(a), we observed a peak structure located at $E – E_F = –1.0$ eV, which is accompanied by another structure at –0.75 eV (Fig. 2(b)) corresponding to the exchange-split $t_{2g}$ bands. Furthermore, we observed bands dispersing between –0.5 eV and $E_F$ corresponding to the $e_g$ bands. A dispersive band with a bottom at –0.5 eV near the Γ point was observed along the in-plane $k_x$ direction (Fig. 2(b)). This band was also observed in a previous ARPES study.[33] Around the X point, we found that the higher intensity near $E_F$ consists of several bands as indicated by polarization-dependent measurements (Figs. 2 (c) and (d)). The band with a bottom around the Γ point, which forms FS named as α, crosses $E_F$ around X (Figs. 2(c) and (d), circularly polarized and s-polarized). Although the intensity derived from a hole band, which forms FS named as β, is quite small for the circularly polarized light, it becomes clearer for the $p$-polarized light, as shown in the right panel in Figs. 2(c) and (d), because of an orbital selectivity of photoexcitation [37,38]. In addition, we found heavy bands located just below $E_F$, which we call γ and δ. Based on these results, we summarize the band structure in Fig. 2(e) and Fig. 2(f). Along the ΓX line, the α, β, and γ bands cross $E_F$ and constitute the FSs. Comparing our experimental electronic structure with the published theoretical band structure [27], the β band is assigned to a majority spin hole band forming a hole FS around the Γ point. The α band also corresponds to a majority spin band around the Γ point. The experimental α band, however, crosses $E_F$ although the corresponding theoretical band does not cross $E_F$. Furthermore, there is seemingly no theoretical band in the occupied states to which we can assign the γ and δ bands. In order to clarify the character of the γ and δ bands, we investigated the spin-resolved band structure by high-resolution SARPES, as described below.

**B. Spin-resolved electronic structure**

Our high-resolution SARPES results are summarized in Fig. 3. Using $hv = 6.994$ eV, the measured $k_x$ line is in the ΓXMX' plane, as shown in Fig. 3(a). We observed an ellipsoidal FS around the Γ point and a linear FS along the $k_y$ direction at $k_x = 0.33$ Å$^{-1}$ as shown in Fig. 3(b), which corresponds to γ and β, respectively, defined in Fig. 1(e).

Figures 3(c)-(h) show SARPES intensity maps taken along the blue line in Fig. 3(b). We observed bands located at $E - E_F = -1$ eV, $-0.5$ eV, $-50$ meV, just below $E_F$, and a dispersive hole band corresponding to the β band. From Fig. 3(i), the bands located at $-1$ eV split into majority (lower energy) and minority (higher energy) spin bands, as marked by black solid lines. These spectral features strongly support the picture shown by GGA calculations that the $t_{2g}$ bands split into the majority and minority spin bands.[10]

For the $e_g$ bands, we obtained the spin-polarized band structure more clearly. From the correspondence to the band structure obtained by our ARPES measurements, we identified the $e_g$ bands as shown in Figs. 3 (c) and (f). At $E - E_F = -0.5$ eV, the SARPES intensity map for the majority spin character has a clear structure corresponding to the bottom of the α band shown in Fig. 2 (c), although the map for the minority spin bands has no structure at $-0.5$ eV. In the same way, it is ascertained that the β band has majority spin character. The result that the spin character of the α and β bands corresponds to the majority spin is consistent with the band structures calculated within GGA.[10] On the other hand, the γ and δ bands have minority spin character, as clearly shown in Fig. 3(e). We discuss detailed characteristics and interpretation of the minority spin bands below.

Here, we discuss the half-metallicity of $CoS_2$. As bands forming FSs, we observed not only the majority spin bands but also the minority spin bands, as shown in Figs. 3 (c)–(h). The minority spin bands are occupied only down 100 meV binding energy, providing spectroscopic evidence for near rather than true half-metallicity of $CoS_2$. The minority spin bands stay down to 20 K as shown later in Fig. 6(a). This result is consistent with claims by Andreev reflection and photoemission studies.[29,31,32] The non-half-metallicity is also supported by the fact that the ordered moment $\mu_s \sim 0.9$ $\mu_B$/Co at 2 K[23,24] is not exactly $\mu_s = 1$ $\mu_B$/Co which is the expected ordered moment when the $e_g$ electrons are completely spin-polarized. Our results are also consistent with the dHvA results which show a lot of dHvA branches derived from the cross sections of FSs in the $k_x$-$k_y$ planes.[25]

As seen in Fig. 3(j), the γ band has a very narrow peak just below $E_F$. Such a very narrow peak just below $E_F$ was observed by high-resolution *spin-integrated* photoemission spectroscopy using the Xe Iα line below $T_C$ [32]. Sato *et al*. interpreted the peak structure as evidence for the occupied minority spin state around the R point, supported by their band-structure calculation results. Our high-resolution SARPES results, however, show that the peak structure is observed at $k_F$ of the minority spin hole band around the Γ point. Note that the minority spin hole band around the Γ point cannot be attributed to a surface band because the band shows dispersion with the periodicity of the bulk Brillouin zone along the $k_z$ direction as shown as the intense band near $E_F$ (see

Supplemental Materials). In addition, a previous study on the surface of $CoS_2(100)$ has shown that the surface bands are located between $E - E_F = -0.7$ eV and $-0.2$ eV along the ΓX direction [39], which supports that the δ band is not to be interpreted as a surface band. For understanding the minority spin bands around the Γ point, we analyze a more detailed comparison of our experimental spectra with the theoretical band structure, as discussed below.

**C. Spin-dependent electron correlation**

Figure 4 (a) shows the theoretical band structure obtained from GGA with spin-orbit coupling (GGA+SO) calculations with exchange reduced to 74%. The theoretical band structure without the exchange reduction is shown in Supplemental Materials. The $E_F$ position is determined to conserve the number of electrons. For the exchange-reduced band structure, the total moment is estimated to be 0.83 $\mu_B$/Co, which is in good agreement with the experimental value [36]. From the correspondence between the theoretical and experimental total moment, it is appropriate to compare our ARPES and SARPES data with the exchange-reduced band structure. Note that even though spin-orbit coupling leads to small additional band splittings, we discuss orbital character in terms of projection on $l$, $m$, $s$ quantum numbers which is still useful.

For the majority spin bands, the overall band structure calculated within GGA+SO is in line with that obtained by ARPES shown in Fig. 2. For example, along the ΓX line, the α and β bands correspond to a band with a bottom at $E - E_F \sim -0.55$ eV at the Γ point and correspond to a light hole band crossing $E_F$, respectively. Furthermore, the orbital character of the α band is the $d_{z^2}$ orbital while the β band is mainly characterized by $d_{x^2-y^2}$ orbital character, based on the correspondence with the theoretical band structure. Note that due to the distortion of the $CoS_6$ octahedra, these two $e_g$ orbitals are not degenerate. The difference of the orbital characters causes the polarization dependence of the ARPES intensity maps obtained around the X point shown in Figs. 2 (c) and (d). For the minority spin bands, the γ band corresponds to a minority spin band whose bottom touches $E_F$ in the middle of the ΓX line in the exchange-reduced band structure.

In order to further discuss the correspondence between the experimental spectra and the theoretical band structures, we compare the band structures and shapes of the FSs obtained by ARPES and by the exchange-reduced theoretical band structure. As shown in Figs. 4(c) and 4(d), the bands α, β, γ, and δ obtained from ARPES measurements can be assigned to the bands obtained from the GGA calculations. As shown in Fig. 4(e), we observed five FSs: α, β, γ, a pocket around the X' point, and a pocket around the M point, in the cut P1, as mentioned above. In Fig. 4(f), we show the theoretical energy contour lines. The energy contour lines were assigned to the experimentally observed FSs based on correspondence of characteristics of the theoretical band structures with those of the experimental band structures; for instance, both experimental contour line and theoretical one which are assigned to FS C are produced by a minority spin hole band. Comparing Fig. 4(e) with Fig. 4(f), square-like contour lines corresponding to β and trapezoidal contour lines corresponding to the pocket around the X' point have characteristic shapes corresponding to those observed by ARPES. This agreement of the theoretical and experimental FSs indicates that the overall electronic structure can be roughly understood by the GGA calculation.

However, there is a deviation of our experimental band structure from our theoretical one in terms of effective mass. Figure 5(a)–(d) show comparisons of the ARPES intensity maps in the ΓX line and the band structure calculated within GGA. The band widths of the experimentally observed majority spin bands (α, β) are well explained by the GGA calculations, while the band width of the γ band observed with 7-eV laser shown in Fig. 5(c) is much smaller than that obtained from the GGA calculations shown in Fig. 5(d). Estimating the effective mass of the γ band from quadratic curve fitting, the experimental effective mass was found to be more than five times larger than that of the GGA calculations. This suggests the magnitude of electronic correlation in $CoS_2$ is strongly dependent on the spin character.

To rigorously discuss the spin dependence of electron correlations, bands with identical orbital character but different spin characters must be compared, because the orbital character-dependent contribution of electron correlations must be eliminated. We show the ARPES intensity map along the MR line overlaid with band plots showing the peak positions, and show the theoretical bands in Figs. 5(e) and (f), respectively. The ε and ζ bands have the same orbital character, mainly $d_{x^2-y^2}$, but have opposite spin. In the theoretical band structure, the ε and ζ bands split at the R point due to spin-orbit coupling, which was not observed by ARPES in which the energy resolution was 20 meV. Focusing on the effective mass, the slope at $k_F$ of the observed ε band is much larger than that of the calculated ε' and ε'' bands. This deviation can be attributed to mass renormalization of the theoretical ε' and ε'' bands due to a correlation effect. On the other hand, the band width of the ζ band is larger than that of the ζ' and ζ'' bands. From the point of view of a tight binding model, a wider bandwidth indicates a smaller effective mass. This suggests that the correlation of the minority spin electrons is larger than that of the majority spin electrons.

Here, we quantitatively compare the magnitude of the spin-dependent electronic correlation of $CoS_2$ with that of typical ferromagnets. For the electron bands around the Γ point of Ni (110), $\eta_\uparrow/\eta_\downarrow$ was estimated to be 1.6, where $\eta_\sigma$ is a mass enhancement factor defined by $\eta_\sigma = m_\sigma^*/m_b$ (index $\sigma$ indicates majority (↑) or minority (↓) spin state) where $m_\sigma^*$ and $m_b$ are effective masses given by ARPES and by GGA calculation, respectively.[20] For the exchange-split electron bands around the R point of $CoS_2$ (namely, the ε and ζ bands), $\eta_\uparrow/\eta_\downarrow$ was estimated to be 1.0/2.6 = 0.38, where we assume that the effective mass of the ζ band is inversely proportional to the bandwidth.

These quantitative estimates are summarized in Table I. From Table I, we make two observations: (i) The mass renormalization of the minority spin band in $CoS_2$ is larger than that of the majority spin band, while in Ni the relation of the mass renormalizations is inverse. (ii) The ratio of $\eta_\uparrow$ and $\eta_\downarrow$ of $CoS_2$, $(\eta_\uparrow/\eta_\downarrow)^{-1} = 2.6$, is much larger than that of Ni(110), 1.6. This leads us to suspect that the near-$E_F$ electronic structure constituted by unrenormalized majority spin bands and the large mass renormalization of minority spin bands are one of characteristic features of a nearly HMF. A study on the half-metallic ferromagnet $CrO_2$ reported a similar result that the majority spin $t_{2g}$ bands can be well reproduced by LSDA calculation with $U_{eff} < 1$ eV, although minority spin bands cannot be captured by the calculations because no minority spin $t_{2g}$ band is observed below $E_F$ for $CrO_2$.[16,19] However, it has been reported that minority spin bands of $CrO_2$ are broadened by increasing thermal fluctuations due to a correlation effect.[18] This suggests strong correlation for the minority spin electrons, similar to the case of $CoS_2$.

In a previous study of $CoS_2$ using soft X-ray ARPES (SXARPES), the large spin-dependent electron correlation as observed in our study was not mentioned [33]. Along the XR direction, the $k_F$ of the ε band is estimated to be ~$0.2\pi/a$ away from the R point in the previous study, while that in our study is estimated to be $0.5\pi/a$ as shown in Fig. S6 in Supplemental Materials. A reason for the observed smaller $k_F$ in SXARPES is due to the difference in the magnitude of the exchange splitting. In the SXARPES results, the bottom of the band corresponding to the ζ band is approximately –0.65 eV, while in our ARPES results it is estimated to be –0.6 eV. Since this difference can be attributed to the difference in the magnitude of the exchange splitting, the energy of the bottom of the ε band is on the high binding energy side with an energy scale comparable to that of the energy difference. Therefore, a larger $k_F$ was obtained in our ARPES compared to the $k_F$ estimated in the SXARPES study. It is still unclear why this difference in exchange splitting occurred, but one possibility is that it was due to sample dependency; calculations by Wu *et al.* reported that a slight shift in the position of S can change the magnitude of the exchange splitting [40].

Based on our SARPES and band calculation results, the effective mass of the observed minority spin bands along the MR line is renormalized significantly, while that of the observed majority spin bands is rather smaller than that estimated from our GGA calculation. Such mass renormalization only for the minority spin electrons has been investigated as spin-dependent electronic correlation in studies on typical ferromagnetic metals.[20–22] One of the studies implies that the imaginary part of the self-energy Σ of Co 3$d$ bands indicates a spin dependence.[22] There, although the electronic states of Co obtained by SARPES measurements cannot be explained by a LSDA+$U$ calculation, which is similar to our findings, their spectra calculated by LSDA+DMFT within the 1-step model improve the description of the experimental photoemission data. This suggests that the deviation of our experimental results from our theoretical band structure can be understood by introducing DMFT.

Note that the spin-dependent electronic correlation is different from an orbital-dependent correlation discussed for iron-based superconductors. For the case of $K_x Fe_{2-y}Se_2$, the magnitude of electronic correlations depends on the orbital character of Fe 3$d$.[41] For the case of the $e_g$ bands in $CoS_2$, the magnitude of electronic correlations depends on the spin, because the band shape of the calculated $e_g^\uparrow$ and $e_g^\downarrow$ bands is the same, and splitting them by an exchange interaction does not change the orbital character.

### D. Dynamical mean-field theory investigation

We performed DFT+DMFT calculations to investigate the spin-dependent renormalization beyond GGA. Technical details are summarized in Appendix C. Figure 7 shows the temperature dependence of the ferromagnetic moment $M$ and the inverse of the ferromagnetic susceptibility, 1/χ, for $J$ = 1.0 eV and $U$ = 4 eV. From the linear fitting of 1/χ, $T_C$ was determined to be 401 K, which is larger than the experimental value of 122 K. If we take into account that DFT+DMFT calculations with density-density interactions tend to overestimate $T_C$,[42] we can consider this parameter set a reasonable choice. Also, the magnetic moment at a low temperature far below $T_C$ was 0.90$\mu_B$/Co which is in good agreement with the experimental value, 0.93$\mu_B$/Co [25]. Numerical results for different parameter sets are provided in the Supplemental Material.

Figure 6 shows the single-particle excitation spectrum calculated within DFT+DMFT for $T/T_C$ = 0.58. In Figs. 6(a) and 6(d), the $t_{2g}$ bands split into majority spin bands and minority spin bands, although the magnitude of splitting energy is not as large as that calculated within DFT+GGA[10]. The top of the majority spin $t_{2g}$ bands has the almost same energy of that of the minority spin $t_{2g}$ bands, while the bottom of the majority spin $t_{2g}$ bands is approximately 0.3 eV lower than that of the minority spin $t_{2g}$ bands. This difference of band width produces finite spin polarization in the density of states (DOS) of the $t_{2g}$ bands. The spin polarization was observed by our SARPES measurements as an energy splitting of the majority spin EDC and minority spin one, as shown in Fig. 3(i). While energy splitting of the $t_{2g}$ bands estimated by GGA calculations is too large to explain the spin splitting observed by SARPES, DFT+DMFT calculations improve the description of the spin splitting of the $t_{2g}$ bands.

As for the $e_g$ bands, the correspondence between experimental and theoretical spectral functions has been improved. Along the ΓX line, two majority $e_g$ bands cross $E_F$ in the theoretical spectral functions: a large hole band around the Γ point and a small hole band near the X point. These bands produce two FSs closed around the Γ point with the majority spin character in the ΓXMX' plane, as shown in Fig. 6(b). For the minority spin character, as shown in Fig. 6(e), unsharp trapezoidal FSs near the X point are also reproduced. In the X"MRM plane, the theoretical calculations show that a closed FS is located around the X" point, as shown in Fig. 6(c). This FS corresponds to the circular FS observed by ARPES around the X" point shown in Fig. 1(f), although the FS does not appear from GGA calculations. (Fig. 4(d)). For the minority spin states, the ε band touches $E_F$ at the R point in the theorical spectral function, which produces an electron pocket as shown in Fig. 6(f). In our ARPES intensity maps of FS shown in Fig. 1(f), intensity at the R point is not so large, because the bottom energy of the ε band is lower than that calculated within DFT+DMFT.

The DFT+DMFT calculations have improved the theoretical description of spin-dependent electron correlation. We compare the band dispersions along the MR line calculated within DFT+DMFT with those observed by ARPES as shown in Fig. 8. In DFT+DMFT, the bandwidths of both ε and ζ bands are reduced compared to those of the GGA calculations. This can be interpreted as a result of the increase in the effective mass due to the electronic correlation effects. Focusing on the magnitude of the bandwidth reduction compared to the GGA calculations, the bandwidth of the ζ band is reduced by 20%, while for the ε band it is reduced by 45%. This spin-dependent reduction of the bandwidth is a result which better explains the spin-dependent electron correlation effects observed in the spectral function obtained by ARPES.

Although the DFT+DMFT yields considerable improvement on the spin-dependent correlations, there are still quantitative deviations from the ARPES results. Figure 8 indicates that the ε band is in quite good agreement, while the bandwidth is underestimated by 0.1 eV for the ζ band. Besides, the mass enhancement factor $\eta_\sigma \equiv m_\sigma^*/m_b$ for the minority spin band is estimated to be 1.7 in DFT+DMFT, while it is 2.6 in the experiment as summarized in Table I. This deviation indicates that DFT+DMFT underestimates the correlation effects. This implies that correlations beyond DMFT play a role in the spin-dependent renormalization in $CoS_2$, which will be discussed in Sec. II.F.

**E. Temperature dependence**

In Fig. 9, we examine the temperature evolution of the ARPES spectra. In Fig. 9 (a), the spectral intensity of the δ band gets broader rapidly with increasing temperature up to $T_C$ (= 122 K). Concomitantly, the β and γ bands are relatively enhanced. Except for the broadening, the band structure does not change significantly with increasing temperature. Moreover, from the energy distribution curves (EDCs) shown in Fig. 9 (b) and (c), although the peak structure at $E_F$ gets broader rapidly as seen in the ARPES intensity map, the peak structure does not disappear even above $T_C$. From wide spectra corresponding to the Co 3$d$ ($t_{2g}$ and $e_g$) bands shown in Fig. 9 (d), a peak structure corresponding to the α band shifts rapidly by +0.16 eV through $T_C$. Note that temperature cycle of the measurements was conducted to make sure that it is an intrinsic behavior, not because of the surface aging. Based on our SARPES results which demonstrate that the α and β (γ and δ) bands are the $e_g^\uparrow$ ($e_g^\downarrow$) bands, the temperature dependence of the ARPES spectra demonstrates that the $e_g^\uparrow$ and $e_g^\downarrow$ bands are not degenerate even above $T_C$. This suggests that the origin of the demagnetization of $CoS_2$ is not Stoner damping but vanishing of the ferromagnetic long-range order by spin fluctuations, as in the case of iron.[43,44]

In a previous spin- and angle-integrated photoemission study, the broadening of the near-$E_F$ peak was interpreted as a change from the ferromagnetic electronic structure to the paramagnetic one, that is, as damping of the exchange splitting.[32] However, by our spin- and angle-resolved measurements, we identified that the exchange splitting remains above $T_C$, although the magnitude of the splitting reduced slightly. Therefore, the broadening of the δ band is rather attributed to a reduction of lifetime of the minority spin electrons. Figure 9 (e) shows the temperature evolution of the width of the spectral peak at $E_F$ ($\Delta E$) shown in (c). $\Delta E$ shows the $T^2$ dependence below $T_C$ but deviates from the $T^2$ curve above $T_C$. The temperature evolution is nicely in line with that of the resistivity.[25] The fact that deviation from the $T^2$ dependence of $\Delta E$ occurs at $T_C$ of the bulk $CoS_2$ sample is consistent with the scenario that the δ band is the bulk state band. Life time of the minority spin electrons ($\tau$) corresponds to $1/\Delta E$ which rapidly increases below $T_C$, as shown in Fig. 9(f). Because the rapid broadening is characterized by $T_C$, it is found that scattering by magnetic fluctuations is a dominant factor contributing to the broadening of the near-$E_F$ peak and the increase of the resistivity. Furthermore, the rapid broadening is significant for the δ band which is highly mass renormalized by correlations. This suggests that electron–electron scattering due to Coulomb interaction is also an indispensable factor contributing to the temperature dependence of the resistivity.

Here, we discuss prospects for realizing a half-metallic electronic structure at room temperature. The thermal broadening is quite similar to that predicted by DMFT calculations for HMFs.[13,45–48] DMFT calculations have demonstrated that by thermal fluctuations the spectrum of the electronic states for either one of spin states forming a gap at $E_F$ is broadened and has a finite value at $E_F$. This state emerging near $E_F$ as a result of thermal fluctuations is called a non-quasiparticle (NQP) state, which has been actually observed experimentally as breaking of the half-metallicity in $CrO_2$ and $Co_2MnSi$.[17,18,48] These facts suggest that although spin-dependent correlations may be a common characteristic of HMFs, as discussed above, electronic correlations break the half-metallicity at finite temperature, unfortunately. Furthermore, our results demonstrate that the $e_g^\uparrow$ bands are well-described by a one-electron picture despite the $3d$ nature, which can be another remarkable feature specific to HMFs observed in $CrO_2$ as well [19]. This indicates that realizing weak correlation in the majority spin electronic states might lead to discovery of new HMFs. The present study offers the new insight that spin-dependent electronic correlations are one of the most important factors that must be understood further for realizing a half-metallic electronic structure at higher temperatures.

**F. Possible origins of the spin-dependent electronic correlations in CoS$_2$**

The large spin-dependent electronic correlations in $CoS_2$ are also observed in half-metal $CrO_2$, as mentioned above, and may in fact be a correlation effect specific to HMFs. Therefore, the spin-dependent electronic correlations could originate from the difference in number between majority and minority spin electrons near $E_F$. The largest electron correlation term is $U \sum_\alpha n_{\alpha\uparrow} n_{\alpha\downarrow}$ where the subscripts $\alpha = d_{z^2}, d_{x^2-y^2}$ denote the orbital component. This term only contains electron interactions between electrons with the same orbital character. In half-metallic electronic states, the magnitude of DOS near $E_F$ varies greatly depending on the spin. Therefore, diagrammatic contributions to the self-energy will reflect the asymmetry in the DOS of majority and minority carriers. This effect might lead to spin dependent effective masses and could thus be responsible for a spin-dependence of electronic correlations. This model is only relevant for half-metallic electronic states, and might universally apply in HMFs.

Another candidate for the origin of spin dependent electronic correlations could be the difference of magnitude of the $d$–$p$ hybridization. A study of GGA calculations shows that the fraction of orbital components that constitute the DOS near $E_F$ varies with the spin orientation.[10] The majority spin states consist of Co $3d$ and S $3p$ orbitals in comparable proportions, while the minority spin states consist almost entirely of Co $3d$ orbitals. This difference in magnitude of $d$–$p$ hybridization could be observed as intensity difference in our ARPES measurements: As shown in Fig. 2, the intensity of the majority spin band is smaller than that of the minority spin band near $E_F$. Considering that the cross section of S $3p$ electrons is smaller than that of Co $3d$ at this photon energy (see Supplemental Material), it can be inferred that the majority spin $e_g$ bands are more strongly hybridized with the S $3p$ bands. Also from the valence charge density, the wave function of the $e_g^\uparrow$ electrons is extended along the Co-S direction, while that of the $e_g^\downarrow$ electrons is more localized around the Co nucleus. This originates from the difference in number between majority and minority spin electrons. The difference of wave-function extent might be another origin of the spin-dependent electronic correlation in $CoS_2$.

These two scenarios both attribute the spin-dependent correlations to the spin-dependent DOS as a characteristic of HMFs. These processes are taken into account by the DFT+DMFT calculations, though it is difficult to conclude which process is dominant in $CoS_2$. On the other hand, quantitative deviations between theory and experiment discussed in Sec. II.D imply the existence of other factors that are not included in the DFT+DMFT calculations. One possibility is electron-magnon interaction discussed in the context of the non-quasiparticle state [13]. It has been shown by perturbation theory that the electron-magnon exchange interaction affects the low-energy states of the minority spin band, which could enhance the spin-dependent effective mass. DMFT takes only *local* spin fluctuations into account and the feedback from the ferromagnetic magnon is missing. We conclude that the large spin-dependent electronic correlations originate in the spin-dependence of DOS plus nonlocal effects such as electron-magnon interactions.

## V. CONCLUSIONS

We have investigated the electronic structure of pyrite-type $CoS_2$ by synchrotron ARPES and high-resolution SARPES. Along the $\Gamma X$ line, we observed localized bands around $E - E_F = -1$ eV and dispersive bands between $-0.5$ eV and $E_F$. The overall character of the observed bands is in good agreement with the Co $3d$ bands calculated within GGA. In addition, localized bands were observed near $E_F$. From our SARPES measurements, we unambiguously identified that the near-$E_F$ bands had the minority spin character, while the GGA calculations predicts that an occupied minority spin band exists solely around the R point. The deviation of the experimental spectra from the theoretical band structure is roughly explained by reducing the exchange splitting of the $e_g$ band in GGA calculations, demonstrating that the GGA calculations overestimated the exchange splitting of $CoS_2$. However, the modified band structure cannot explain the bandwidth of the minority spin bands, although the bandwidths of the majority spin bands are in line with the theoretical bandwidths. This can be interpreted as a result of a spin-dependent electronic correlation effect. The ratio of the mass enhancement factors $(\eta_\uparrow/\eta_\downarrow)^{-1}$ was estimated to be 2.6 which is much higher than that of a typical ferromagnetic metal like Ni where $\eta_\uparrow/\eta_\downarrow = 1.6$ We performed DFT+DMFT calculations in order to unveil the origin of the large spin-dependent electronic correlations. DFT+DMFT improves the description of the SARPES spectra compared to GGA, but a quantitative deviation still remains. In particular, the large effective mass of the minority spin band was still underestimated by the DFT+DMFT calculations. This result implies the importance of spatial correlations beyond DMFT for the anomalously large spin-dependent electronic correlations in half-metallic materials. From our temperature-dependent ARPES experiments, we observed a decoherence of the mass-renormalized minority spin electrons which is much more rapid than that of the majority spin band. The decoherence of the minority spin band is scaled by $T_C$ of $CoS_2$, indicating that the decoherence originates from increase of scattering of the conducting minority spin electrons by magnetic excitations. From the variation of the $e_g$ bands, the majority and minority spin bands are not damped even above $T_C$, as seen in Fe metal. This indicates that the thermal demagnetization in $CoS_2$ is due to spin fluctuations. Through our thorough electronic-structure study of nearly HMF $CoS_2$, we suggest that spin-dependent electronic correlations are an important factor which must be taken into account not only for understanding phenomena appearing in spin-polarized strongly correlated systems, but also for designing half-metallic materials via theoretical calculations.

**Acknowledgement**

The laser-based SRPES experiments were conducted at ISSP with the approval of ISSP (Proposal Nos. B248, A182, and A177). Part of the ARPES experiments were performed under the Photon Factory Proposal No. 2016G667. This work was partially supported by the Program for Promoting the Enhancement of Research Universities and a Grant-in-Aid for the Japan Society for the Promotion of Science (JSPS) Fellows (Grant No. 16J03208), and JSPS KAKENHI (Grants No. JP18K03484, No. 20K20522, No. 20H01853, No. 20K20522, No. 21H01041, No. 21H01003, No. 21H01041, and No. 21H01003) from the Ministry of Education, Culture, Sports, Science and Technology of Japan (MEXT). H. F. was supported by a Grant-in-Aid for JSPS Fellows. Part of the computations were carried out at the Supercomputer Center at the Institute for Solid State Physics, the University of Tokyo. This work was supported by MEXT Leading Initiative for Excellent Young Researchers. We thank A. Ino for providing his ARPES analysis program.

**APPENDIX A: Samples and photoemission measurements**

Single crystals of $CoS_2$ were synthesized by the vapor transport method,[25] and were used for ARPES and SARPES measurements. ARPES experiments were carried out at beamline BL-28A of the Photon Factory (PF), KEK, using synchrotron radiation as the excitation light source. Circularly polarized light was used, except for the polarization-dependence ARPES measurements. The total energy and angular resolutions were set to approximately 20 meV and 0.3° (corresponding to ~0.02 Å), respectively. $E_F$ of the samples was calibrated by measuring a gold foil that was electrically connected to the samples. The data were taken at $T$ = 50 K. Laser-based ARPES and SARPES experiments were performed at the Institute for Solid State Physics, The University of Tokyo.[49] The $p$-polarized light with $hv$ = 6.994 eV was used to excite the photoelectrons. Photoelectrons were analyzed with a combination of a ScientaOmicron DA30L analyzer and a very-low-energy-electron-diffraction (VLEED) type spin detector. The energy and angular resolutions were set to 6 meV and 0.3° (corresponding to ~0.005 Å) for spin-integrated ARPES and 10 meV and 1° (corresponding to ~0.02 Å) for SARPES, respectively. the base pressure was kept below $1 \times 10^{-8}$ Pa. Calibration of $E_F$ for the samples was achieved using a gold reference. The data were taken at $T$ = 40 K, except for the temperature-dependence data in Fig. 6. Clean surfaces for all measurements were obtained by *in situ* cleaving of the samples. The high structural quality of a sample (100) surface was verified by LEED, as shown in Fig. 1(a). The sample was magnetized along the in-plane [010] direction by bringing the sample close to a magnet at low temperature below $T_C$. The approximate magnitude of the magnetic field at the sample position was 600 Oe.

**APPENDIX B: Data processing**

The experimental FS maps in Fig. 1 were obtained with an energy-integration window of ±0.01 eV, and those in Fig. 3 were done with an energy-integration window of ±1 meV around $E_F$. In order to obtain absolute values of spin polarization using the VLEED detector, we used the equation $P = (1/M_0 S_{eff}) (I_+ - I_-)/(I_+ + I_-)$, where $M_0$ is the ratio of the remanent magnetization of the sample to the saturated magnetization, a calibration factor in terms of magnetic domain, $S_{eff}$ (=0.25) is the effective Sherman function of the apparatus and $I_{+(-)}$ is the intensity of the electrons reflected by the positively (negatively) magnetized target. Then, we obtain the majority ($I_\uparrow$) and minority ($I_\downarrow$) spin spectra using $I_{\uparrow(\downarrow)} = (1 +(-) P)(I_{tot}/2)$, where $I_{tot} = I_+ + I_-$.[50,51]

**APPENDIX C: Theoretical calculations**

We performed density functional theory calculations for $CoS_2$ within the full-potential local orbital (FPLO)[52] basis, using the Perdew-Burke-Ernzerhof parameterization of the generalized gradient approximation (GGA) exchange correlation functional[53]. We based our calculations on the pyrite structure of $CoS_2$ given in Ref. 36 with lattice constant $a = 5.539$ Å and space group $Pa\bar{3}$. $k$ meshes of $12 \times 12 \times 12$ in the first Brillouin zone were used. Using the FPLO projective Wannier functions [54], we constructed a 44-band tight-binding model including Co-3$d$ and S-3$p$ orbitals (the unit cell includes four formula units).

We performed the DMFT self-consistency calculations using DCore [55] implemented on the TRIQS library [56]. Summations over $k$ and $\omega_n$ are performed with $n_k = 8^3$ points and $n_{iw} = 4096$ points (for positive frequencies), respectively. The effective impurity problems were solved with the hybridization-expansion algorithm [57,58] of the continuous-time quantum Monte Carlo method implemented as CT-HYB-SEGMENT [59] on ALPSCore library [60]. The ARPES spectra $A(k,\omega) = \text{Tr}(-1/\pi)\text{Im}\hat{G}(\omega + i0)$ were computed by the analytical continuation from the Matsubara frequency to real frequency using the Padé approximation [61]. More information is available in the Supplemental Material.

Commun. **213**, 235 (2017).

[61] H. J. Vidberg and J. W. Serene, Solving the Eliashberg equations by means of *N*-point Padé approximants, J. Low Temp. Phys. **29**, 179 (1977).

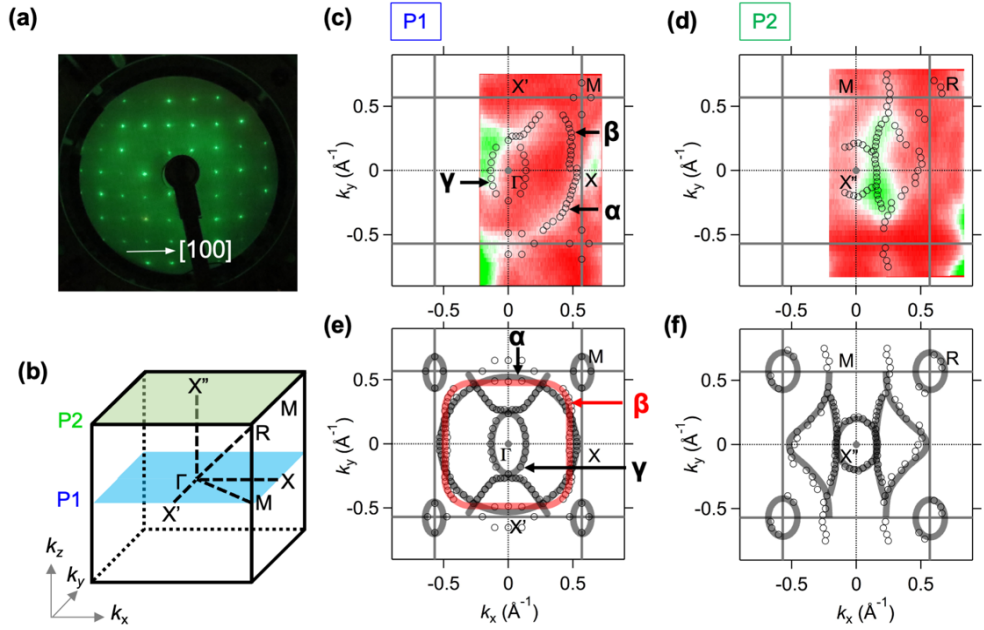

**Fig. 1 (a)** LEED pattern at an incident electron energy of 185 eV. **(b)** Brillouin zone of pyrite-type $CoS_2$. **(c),(d)** ARPES intensity maps of the FS measured at $h\nu = 66$ eV and at $h\nu = 87$ eV (corresponding to cuts P1 and P2 marked in (b)), respectively, overlaid with the fitted MDC peak positions at $E=E_F$ showing the $k_F$ positions (black circles). **(e),(f)** $k_F$ plots along P1 and P2, respectively, symmetrized by assuming a twofold symmetry with respect to the Γ point. The thick shaded lines are guides for the eye.

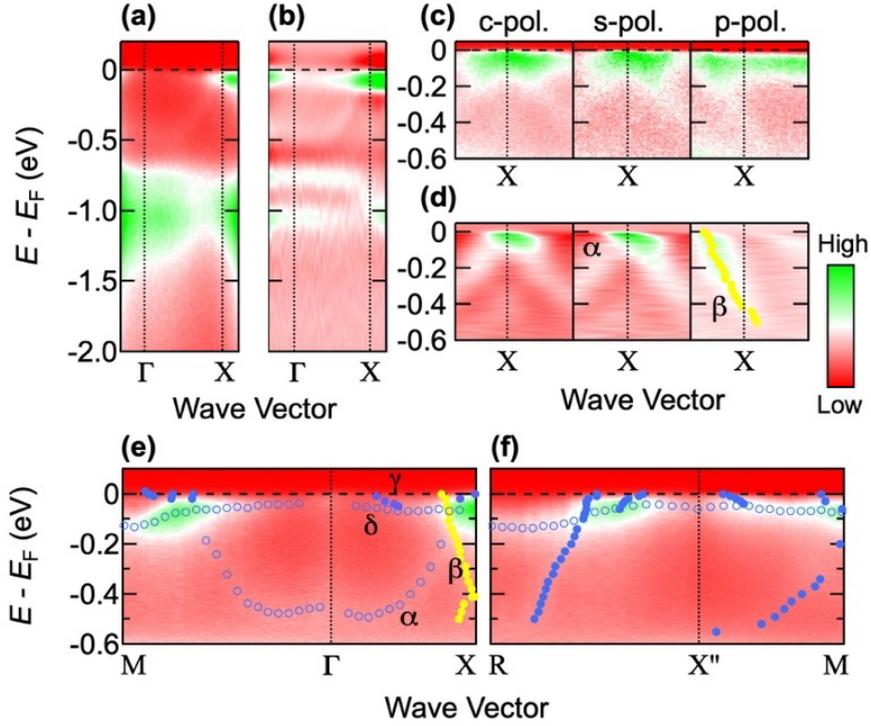

**Fig. 2 (a)** ARPES intensity map as a function of energy up to $E_F$ and wave vector. **(b)** Second derivative of ARPES intensity map in (a) with respect to energy. **(c)** Polarization-dependent ARPES intensity map around the X point: (left) circularly polarized light; (middle) *s*-polarized light; (right) *p*-polarized light. **(d)** Second derivatives of ARPES intensity maps in (c) with respect to wave vector. Yellow filled circles show MDC peak positions estimated from the right panel in (c). **(e),(f)** ARPES intensity maps along the MΓX and the RX"M lines, respectively, overlaid with plots showing peak positions of EDCs (empty circles) and MDCs (filled circles). Yellow filled circles are the same as in (d) but folded with respect to the X point.

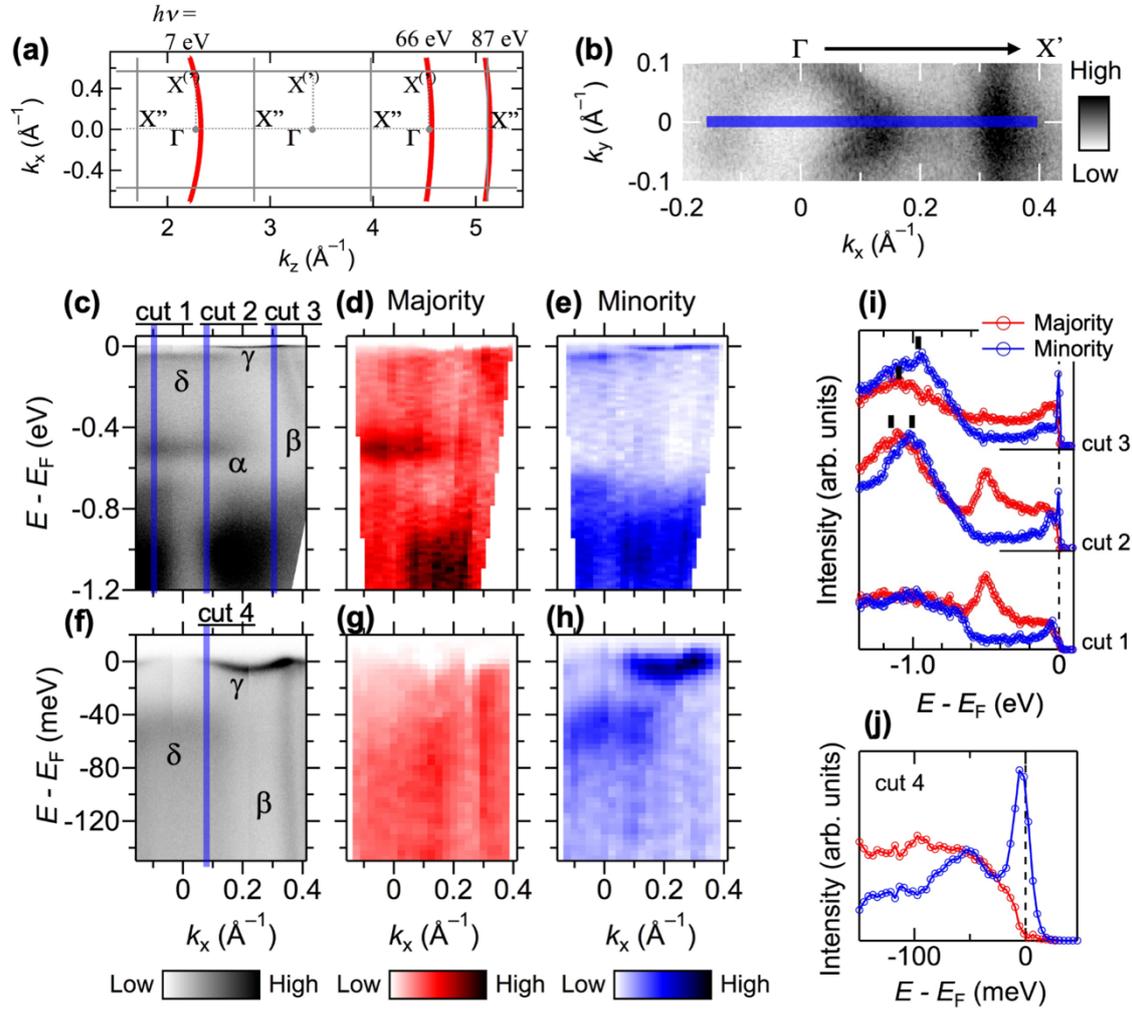

**Fig. 3 (a)** Two-dimensional Brillouin zone (gray lines) showing estimated measured $k$ positions for the states near $E_F$ (red curves) for each photon energy. **(b)** ARPES intensity map of the FS as a function of $k_x$ and $k_y$ measured by 7-eV laser. The FS is obtained with an energy-integration window of ±1 meV. **(c)–(e)** ARPES (c) and SARPES ((d): Majority spin bands; (e) Minority spin bands) intensity maps as a function of energy to $E_F$ and $k_x$. **(f)–(h)** Same as (c)–(e) but measured in the vicinity of $E_F$. **(i),(j)** Spin-resolved EDCs at k points corresponding to the blue shaded line cuts in (c) and (f). Black solid makers in (i) indicate peak positions of the structure around −1 eV in the majority and minority spin spectra.

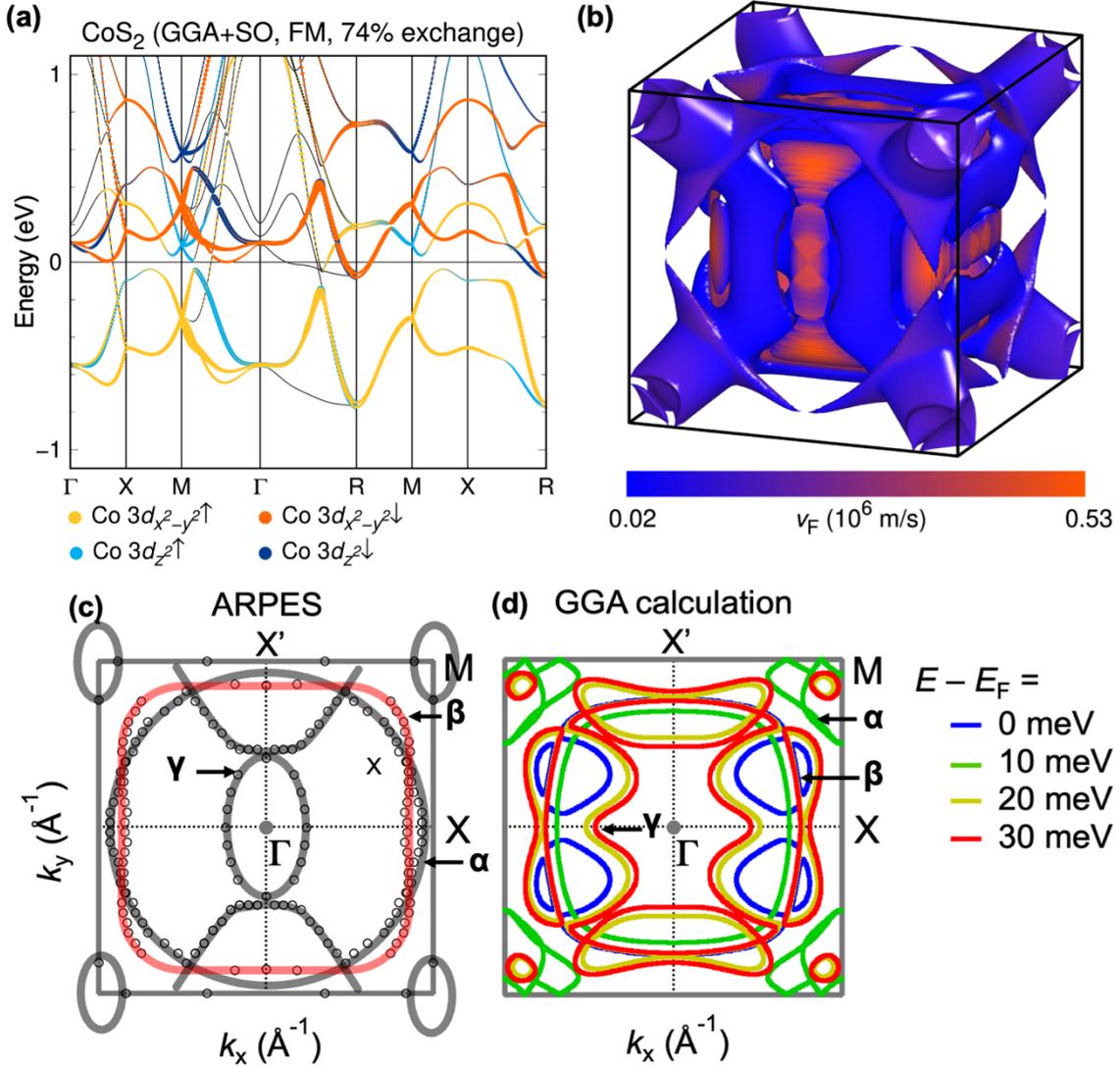

**Fig. 4** Electronic structure of CoS$_2$ in ferromagnetic state (FM) determined with GGA+SO exchange correlation functional. (a) Band structure calculated with exchange splitting reduced to 74%. Both majority and minority bands are crossing $E_F$. (b) Corresponding Fermi surface with additional Fermi surface sheets. (c) Experimentally obtained $k_F$ plots, same as Fig. 1(g). The thick shaded lines are guides for the eye. (d) Theoretical contour plots at $E - E_F = 0$ meV (blue; corresponding to FS), $E - E_F = 10$ meV (green), $E - E_F = 20$ meV (yellow), and $E - E_F = 30$ meV (red), calculated within GGA.

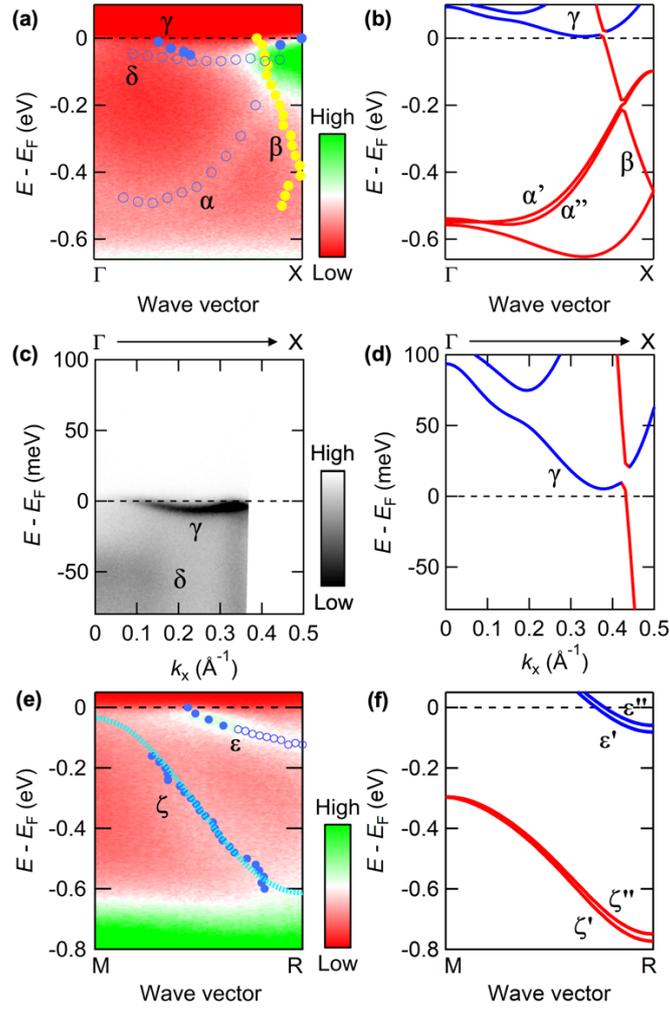

**Fig. 5** Comparison of our experimental ARPES results with band structures calculated within GGA. **(a)** ARPES intensity map taken by synchrotron light along the ΓX direction with peak position s of EDCs (blue empty circles) MDCs (blue and yellow filled circles), same as Fig. 1(e). **(b)** Band structure along the ΓX direction obtained from GGA+SO calculations with exchange splitting reduced to 74%. Red and blue colors indicate the projected majority and minority spin character, respectively. **(c)** Same as (a) but taken by 7-eV laser. **(d)** Same as (b) but enlarged near $E_F$. **(e)** Same as (a) but taken along the MR direction. The light blue dotted line is the curve of the ζ band fitted with a cosine function to estimate the bandwidth. **(f)** Same as (b) but for the MR direction.

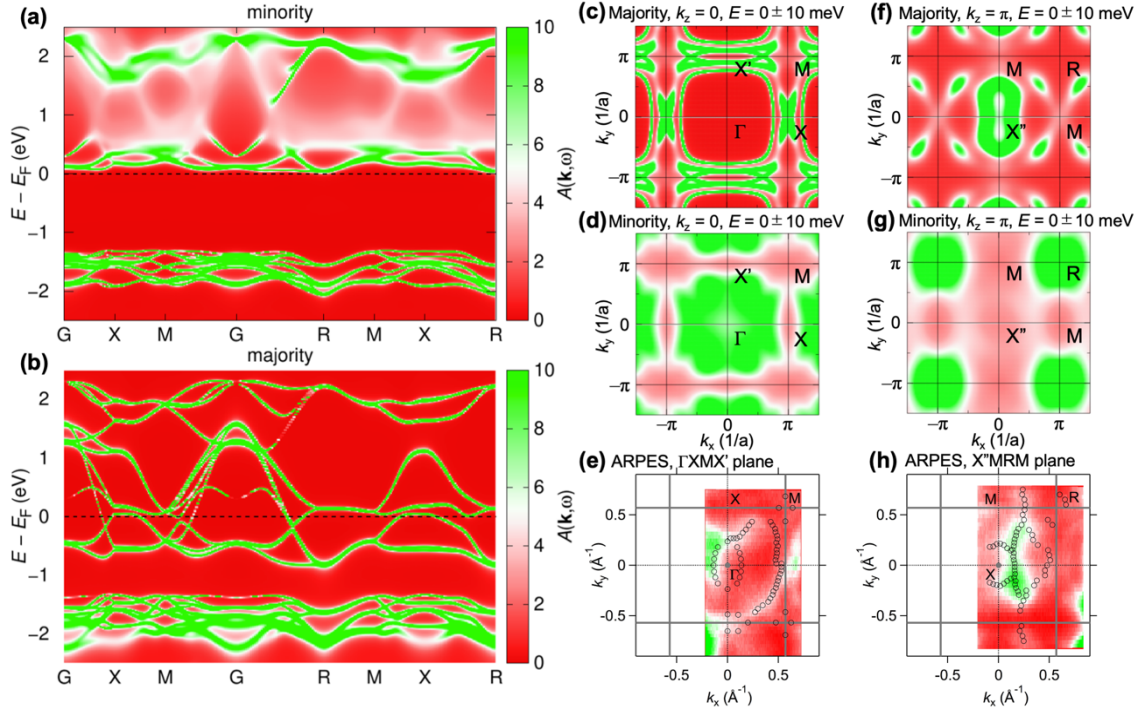

Fig. 6 The single-particle excitation spectrum for $T/T_C = 0.58$, $U = 4$ eV, and $J = 1.0$ eV for the majority spin (a), and for the minority spin (b) calculated within DFT+DMFT. Momentum dependence of the zero-energy spectral functions in the ΓXMX' plane in a 20 meV energy window for the majority spin (c), for the minority spin (d) in calculated within DFT+DMFT, and obtained by ARPES (e). (f)–(h) Same as (f)–(h) but in the X"MRM plane.

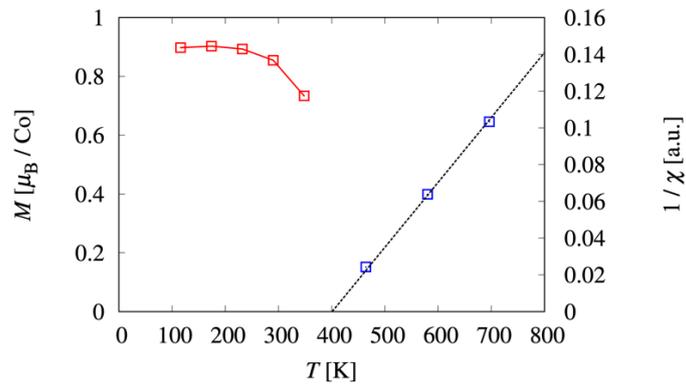

Fig. 7 Magnetic moment *M* (red squares) and inverse of the ferromagnetic susceptibility $1/\chi$ (blue squares) as a function of temperature, obtained DMFT calculations for $U = 4.0$ eV, $J = 1.0$ eV. $T_C$ was determined from a linear fit of $1/\chi$ by the condition $1/\chi = 0$.

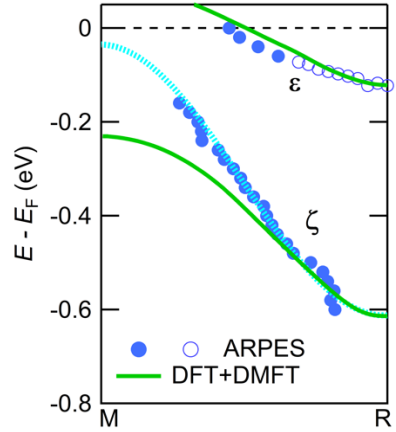

Fig. 8 Comparison of our ARPES results with the spectral function maxima calculated within DFT+DMFT. Blue empty (filled) circles show the peak positions of EDCs (MDCs). Light blue dotted line is the curve of the ζ band fitted with a cosine function. Solid green lines show the peak position of the single-particle excitation spectra calculated within DFT+DMFT, which are shifted so that the energy at the R point matches the experimental value.

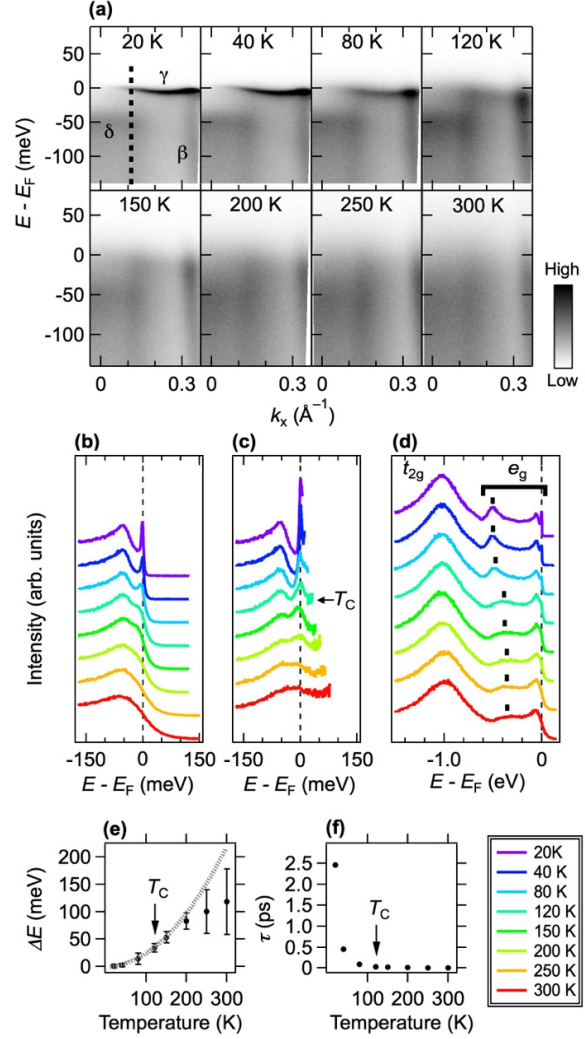

**Fig. 9 (a)** Temperature dependence of the ARPES intensity map along the ΓX line measured by 7-eV laser. **(b),(c)** EDCs at a $k_x$ point marked by the dotted line in (a), and those divided by the Fermi function at measured temperatures which is convoluted with a 6 meV Gaussian corresponding to the energy resolution of the measurement. The EDCs are normalized by their intensity at –150 meV. **(d)** EDCs at a $k_x$ point marked by the dotted line in (a). The EDCs are normalized by the peak height at 1.0 eV. Black thick markers show positions of the structure corresponding to the α band. **(e)** Temperature evolution of the width of the spectral peak at $E_F$ ($\Delta E$) shown in (c). The $\Delta E$ plots are obtained after subtracting the energy resolution of our measurement, 6 meV. The gray dotted line denotes a fit of $\Delta E = A + BT^2$, where $A$ and $B$ are fitting parameters and $T$ is the temperature, to the data in a temperature range of 20–120 K. **(f)** Temperature evolution of the lifetime (τ) of the minority spin electrons occupying the δ band, estimated by $\tau = \hbar/\Delta E$.

Table I. Effective mass ($m^*$) compared with the mass given by the GGA calculation ($m_b$), and the spin-dependent mass renormalization factor which is defined by the ratio between the effective mass for the majority spin ($\eta_\uparrow$) and that for the minority spin ($\eta_\downarrow$).

|  |  | $\eta_\sigma \equiv m_\sigma^* / m_b$ Experiment | $\eta_\uparrow / \eta_\downarrow$ | $\eta_\sigma \equiv m_\sigma^* / m_b$ DFT+DMFT | $\eta_\uparrow / \eta_\downarrow$ |
|---|---|---|---|---|---|
| CoS$_2$ | Majority | 1.0 | 1.0/2.6 = 0.38 | 1.2 | 0.70 |
|  | Minority | 2.6 |  | 1.7 |  |
| Ni [20] | Majority | 2.2 | 1.6 |  |  |
|  | Minority | 1.4 |  |  |  |

# Supplemental Materials for
# Anomalously large spin-dependent electron correlation in "nearly" half-metallic ferromagnet $CoS_2$

Hirokazu Fujiwara[1,2]*, Kensei Terashima[3], Junya Otsuki[3], Nayuta Takemori[3], Harald O. Jeschke[3], Takanori Wakita[3], Yuko Yano[1], Wataru Hosoda[1], Noriyuki Kataoka[1], Atsushi Teruya[4], Masashi Kakihana[4], Masato Hedo[5], Takao Nakama[5], Yoshichika Ōnuki[5], Koichiro Yaji[2], Ayumi Harasawa[2], Kenta Kuroda[2], Shik Shin[2], Koji Horiba[6], Hiroshi Kumigashira[6], Yuji Muraoka[1,3], and Takayoshi Yokoya[1,3]**

[1]*Graduate School of Natural Science and Technology, Okayama University, Okayama 700-8530, Japan*
[2]*Institute for Solid State Physics, The University of Tokyo, Kashiwa, Chiba 277-8581, Japan*
[3]*Research Institute for Interdisciplinary Science, Okayama University, Okayama 700-8530, Japan*
[4]*Graduate School of Engineering and Science, University of the Ryukyus, Nishihara, Okinawa 903-0213, Japan*
[5]*Faculty of Science, University of the Ryukyus, Nishihara, Okinawa 903-0213, Japan*
[6]*Photon Factory, Institute of Materials Structure Science, High Energy Accelerator Research Organization (KEK), 1-1 Oho, Tsukuba 305-0801, Japan*

*Correspondence to: fujiwara.h@s.okayama-u.ac.jp
**Correspondence to: yokoya@cc.okayama-u.ac.jp

**Estimation of inner potential**

In Fig. S1(a), we show the ARPES intensity map showing $k_z$ vs $k_x$ Fermi surface of $CoS_2$. In order to clarify the correspondence relation between the photon energy and the measuring $k_z$ value, we determined the inner potential of $CoS_2$. The perpendicular component of the crystal wave vector $k_z$ of a photoelectron is not conserved across the crystal surface to vacuum interface, but instead can be found using the inner potential correction: $k_z = (2m[(E_i + hv - W)]\cos^2\theta + V_0)^{1/2}/\hbar$, where $m$ is the rest mass of an electron, $E_i$ is the energy of the initial state of the electron, $W$ is the work function, $\theta$ is the emission angle of the photoelectron, $V_0$ is the inner potential.[S1] From the equation, we can determine $V_0$ by observing the periodicity of $E_i$ with respect to $hv$. The observed band structure along the $k_z$ direction shows clearly a high-intensity band located between –0.1 eV and $E_F$. This band is found to produce a hole pocket, as shown by the shaded circle in Fig. S1(a), which suggests that the center of the hole pocket, $k_z \sim 4.55$ Å$^{-1}$, corresponds to a high symmetry point of the Γ or the X" point. In Figs. S1(c) and (d), the structure which is located around $E - E_F = -0.45$ eV is centered around the high symmetry point. This structure corresponds to an electron band with a bottom near the Γ point. The $k_x$–$k_z$ map at $E$–$E_F$ = –0.3 eV shown in Fig. S1(b) also shows that a circular structure is formed around the high symmetry point. Therefore, this high symmetry point can be regarded as the Γ point. The photon energy at which the band near $E_F$ forms a bottom corresponds to the X" point (Fig. S1(c)). Based on the correspondences, we determined the inner potential $V_0$ to be 18 eV. For $V_0$ = 18 eV, the measured $k_x$–$k_y$ planes for the states near $E_F$ using $hv$ = 66 eV and $hv$ = 87 eV are the ΓXMX' plane and the X"MRM plane, respectively.

**Valence band structure**

Fig. S2 shows the ARPES intensity map of the valence band, where the band dispersion of S 3$p$ is observed below $E$–$E_F$ = –2 eV, but its intensity is much small. This suggests that the excitation cross section of S 3$p$ is much smaller than that of Co 3$d$.

**Aging of the γ band**

Figure S3 shows ARPES intensity maps at 20 K measured with 7-eV laser before and after the temperature-dependent measurements. The intensity map did not change before and after the measurement. This indicates that the temperature dependence discussed in the main text is a change in the spectral function associated with the magnetic ordering rather than sample aging. The results also show that the γ band is robust over time, implying that the γ band can be regarded as a bulk-state band.

**Details of the band structure along the MR lines**

Figure S4 shows the raw EDCs and MDCs along a MR line discussed in the main text. Figure S5 shows the ARPES intensity maps of the three MR lines in the Brillouin zone; Figure S5(b) is the same as the data in the text. The majority spin band with a bottom near –0.6 eV (the ζ band in the main text) and the minority spin band with a bottom near –0.1 eV (the ε band in the main text) were observed, as mentioned in the main text. The photoelectron spectrum along the black line in Fig. S5(d) shows three structures: the broadened peak around –0.4 eV corresponding to the ζ band, the Fermi edge corresponding to the ε band, and a peak structure around –0.1 eV. This band around –0.1 eV does not exist in the band calculations of $CoS_2$. The possible origin of this band is either a surface band or the observation of a band in the XM direction due to the $k_z$-broadening effect caused by the surface sensitive measurement. From the $E$-$k$ map in the MR direction (Fig. S5(i)), which is a high-symmetry line in other $k_z$ plane, a band with a top at $k_x = 0$ Å$^{-1}$ is observed near $E_F$. Either way, this band can be regarded as an unrelated band to the spin-dependent electron correlation effect in the main text.

We discuss the three-dimensionality of the ε band by comparing the $E$-$k$ map in the MR direction in the XMRM plane with that in the XM direction in the ΓXMX plane. Figures S5(g–i) show the ARPES intensity maps along the XM direction in the ΓXMX plane overlaid with the ε band plot (blue circles). The electron band around the zone corner, which has the same $k_F$ as the ε band, is not clearly observed along any XM direction in the ΓXMX plane. This suggest that the ε band is a bulk state band.

**Band structure along the XR line**

Figure S6(a) shows ARPES intensity map taken along the X"R direction. From Fig. S6(a), it is found that two bands cross $E_F$, which provide two $k_F$s, $k_{F1}^{Exp}$ and $k_{F2}^{Exp}$. Comparing with the spectral function shown in Fig. S6(b), $k_{F1}^{Exp}$ and $k_{F2}^{Exp}$ can be interpreted as corresponding to the Fermi wavenumbers of the ζ and ε" bands, respectively. On the other hand, comparing with the band structure calculated within GGA, $k_{F1}^{Exp}$ corresponds to the Fermi wavenumber of the ε' band, not that of the ζ band. Although it is difficult to distinguish these two band pictures from our ARPES results, both band pictures show that $k_{F2}^{Exp}$ is the Fermi wavenumber of the ε" band. Comparison of Figs. S6(a) and (c) also shows the large spin-dependent electron correlation along the XR direction.

**Estimation of the mass enhancement factor**

Here, we describe how to estimate the mass enhancement factor. In order to estimate the mass enhancement factor for the majority spin, $\eta_\uparrow$, the bandwidth of the ζ band was determined by fitting with a cosine function. From the ratio of the experimental to theoretical band width, we estimated $\eta_\uparrow$ to be 0.82. However, it is physically inappropriate for $\eta_\uparrow$ to be less than 1, since the effective mass should be increased by Coulomb repulsion. Therefore, in the discussion of the spin-dependent mass renormalization, the value of 1.0 was used as $\eta_\uparrow$. To estimate $\eta_\downarrow$, parabolic curve fittings of the experimental and theoretical ε bands were performed. As the parabolic curve, $(\hbar^2/2m^*)k^2 + E_{\text{bottom}}$ was used, where fitting parameters $m^*$ and $E_{\text{bottom}}$ are the effective mass and the bottom energy of the fitted band. $\eta_\downarrow$ was estimated from the ratio of $m^*$ determined by the fittings of the experimental and theoretical ε bands.

**Setup of GGA calculations**

We performed density functional theory calculations for $CoS_2$ within the full-potential local orbital (FPLO)[S3] basis, using the Perdew-Burke-Ernzerhof parameterization of the generalized gradient approximation (GGA) exchange correlation functional[S4]. We based our calculations on the pyrite structure of $CoS_2$ given in Ref. S7 with lattice constant $a$ = 5.539 Å and space group $Pa\bar{3}$. $k$ meshes of 12 × 12 × 12 in the first Brillouin zone was used. Using the FPLO projective Wannier functions[S5], we constructed a 44-band tight-binding model including Co-3$d$ and S-3$p$ orbitals (the unit cell includes four formula units).

Figure S7(a) shows the band structure of $CoS_2$ in the paramagnetic phase obtained by GGA calculations and its fit by the projective Wannier functions. It shows a perfect fitting of the band structure between –8eV and +3eV. The number of electrons within the effective 44-band model is 68 per unit cell. Figures S7(b) and (c) show the band structure and Fermi surface of $CoS_2$ in the ferromagnetic phase.

**Details of DMFT calculations**

We evaluated correlation effect on the ARPES spectra using the dynamical mean-field theory (DMFT)[S6,S7]. The Matsubara Green's function of the 44-band model is given by

$$\hat{G}_\sigma(i\omega_n) = \left[i\omega_n + \mu - \hat{H}_{\text{DFT}}(k) + \hat{\Sigma}_{\text{dc}} - \hat{\Sigma}_{\text{loc}}(i\omega_n)\right]^{-1} \quad (S1)$$

where the quantities with hat are (44 × 44)-matrices for the band indices. $\hat{H}_{\text{DFT}}(k)$ is the one-body Hamiltonian constructed by DFt. $\hat{\Sigma}_{\text{dc}}$ is a double-counting correction. We estimated $\hat{\Sigma}_{\text{dc}}$ by the Hartree-Fock approximation using the bare Green's function, $\hat{G}_{0,\sigma}(i\omega_n) = \left[i\omega_n + \mu - \hat{H}(k)\right]^{-1}$. $\hat{\Sigma}_{\text{loc}}(i\omega_n)$ is the local (momentum-independent) self-energy to be computed by the DMFT.

We consider correlations only between electrons on the $e_g$ orbitals. The local self-energy $\hat{\Sigma}_{\text{loc}}(i\omega_n)$ then has non-zero matrix elements on the diagonals that represent the $e_g$ orbitals on each Co atom. We represent these diagonal components of the self-energy as, $\Sigma^i_{\alpha\sigma}(i\omega_n)$, where the subscripts $\alpha = d_{z^2}, d_{x^2-y^2}$ and $\sigma = \uparrow, \downarrow$ denote the orbital and spin components, respectively. The superscript $i$ = 1,2,3,4 is an index for four Co atoms. Actually, the four Co atoms are symmetry equivalent, and we omit the index $i$ as $\Sigma^i_{\alpha\sigma}(i\omega_n) = \Sigma_{\alpha\sigma}(i\omega_n)$.

The local self-energy is computed by solving a single-impurity model. We consider density-density terms in the Kanamori-type interaction, which is given by

$$\mathcal{H}_{\text{int}} = U \sum_\alpha n_{\alpha\uparrow} n_{\alpha\downarrow} + U' \sum_\sigma n_{1\sigma} n_{2\bar{\sigma}} + (U' - J) \sum_\sigma n_{1\sigma} n_{2\sigma} \quad (S2)$$

where $\bar{\sigma}$ stands for the spin component opposite to σ. The inter-orbital Coulomb interaction $U'$ is determined from $U$ and $J$ by $U' = U - 2J$ as usual. The intra-orbital Coulomb interaction $U$ is fixed at $U$ = 4. The value of $J$ will be determined so that the calculated ferromagnetic transition temperature is consistent with the experiment.

We performed the DMFT self-consistency calculations using DCore[S8] implemented on TRIQS library[S9]. Summations over $k$ and $\omega_n$ are performed with $n_k = 8^3$ points and $n_{iw}$ = 4096 points (for positive frequencies), respectively. The effective impurity problems were solved with the hybridization-expansion algorithm[S10,S11] of the continuous-time quantum Monte Carlo method implemented as CT-HYB-SEGMENT[S12] on ALPSCore library[S13]. The ARPES spectra $A(k,\omega) = \text{Tr}(-1/\pi)\text{Im}\hat{G}(\omega + i0)$ was computed by the analytical continuation from the Matsubara frequency to real frequency using the Padé approximation[S14].

In order to estimate the ferromagnetic transition temperature, we computed the momentum dependent susceptibility. More generally, we investigated possibility of not only the magnetic transition but also orbital ordering by evaluating general susceptibilities representing fluctuations of the operator $O^i_{\alpha\sigma,\alpha'\sigma'} \equiv c^\dagger_{i\alpha\sigma}c_{i\alpha'\sigma'}$. We solved the Bethe-Salpeter equation to obtain (64 × 64) susceptibility matrix $\hat{\chi}(q)$, and then diagonalize the matrix to obtain 64 eigenvalues[S15]. We confirmed that the largest eigenvalue is located at $q = 0$ and its eigenvector corresponds to the fluctuations of $O_{\text{FM}} \equiv (1/4)\Sigma_{i\alpha\sigma}\sigma O^i_{\alpha\sigma,\alpha'\sigma'}$, namely, the ferromagnetic fluctuation. We thus represent the leading eigenvalue of $\hat{\chi}(q = 0)$ as $\chi_{\text{FM}}$.

**Ferromagnetic moment and the transition temperature**

We obtained a ferromagnetic solution at low temperatures. Figure S8 shows the ferromagnetic moment $M$ as a function of $T$ for several values of $J$. The value of $M$ at a low enough temperature ranges from $M \simeq 0.86\mu_B$/Co for $J = 0.9$ to $M \simeq 0.93\mu_B$/Co for $J = 1.2$.

The transition temperature was determined from the divergence of the ferromagnetic susceptibility. Figure S9 shows the T dependence of the inverse of the ferromagnetic susceptibility $\chi_{\text{FM}}$ in the paramagnetic state. We determined $T_C$ by $1/\chi_{\text{FM}} = 0$. The negative susceptibility for $T < T_C$ indicates that the paramagnetic solution is not the true solution.

Thus obtained $T_C$ is shown in Fig. S10 as a function of $J$. The ferromagnetic solution appears above $J = J_c \approx 0.85$, and $T_C$ monotonically increases with $J$. The transition temperature in the real material is $T_C^{\text{exp}} = 122$ K. Considering the fact that the DFT+DMFT calculations with the density-density interactions tend to overestimate $T_C$ by a factor of 2[S16], we conclude that the reasonable value of $J$ is between 0.9 and 1.0. We use $J = 1.0$ in the main text and in the following calculations. The transition temperature for $J = 1.0$ is $T_C = 401$ K.

**Renormalization factor**

Figure S11 shows the spin- and orbital-dependent renormalization factor $Z_{\alpha\sigma}$ evaluated by $Z_{\alpha\sigma} = [1-\text{Im}\Sigma_{\alpha\sigma}(i\omega_0)/\omega_0]^{-1}$, where $\omega_0 = \pi T$. The orbital dependence is small in both $T > T_C$ and $T < T_C$, we omit the orbital index α in $Z_{\alpha\sigma}$ and discuss only the spin dependence.

In the paramagnetic state, the renormalization factor reaches $Z \approx 0.3$. This explains that the spectrum for $T > T_C$ is broadened over the whole energy region. $Z_{\alpha\sigma}$ splits into two below $T_C$ and the renormalization becomes weaker than in the paramagnetic phase. The spin-up component (minority spin) is more renormalized than the spin-down component (majority spin), namely, $Z_\uparrow < Z_\downarrow$. The ratio of the effective mass between spin up and spin down is estimated as $m^*_\uparrow/m^*_\downarrow = (Z_\uparrow/Z_\downarrow)^{-1} \approx (0.58/0.83)^{-1} \approx 1.4$ at $T/T_C = 0.43$ ($T = 174$ K).

**ARPES spectra**

Figures S12–S16 show the spin- and angle-resolved single-particle excitation spectrum $A(k,\omega)$ for temperatures between $T/T_C = 1.15$ and $T/T_C = 0.43$.

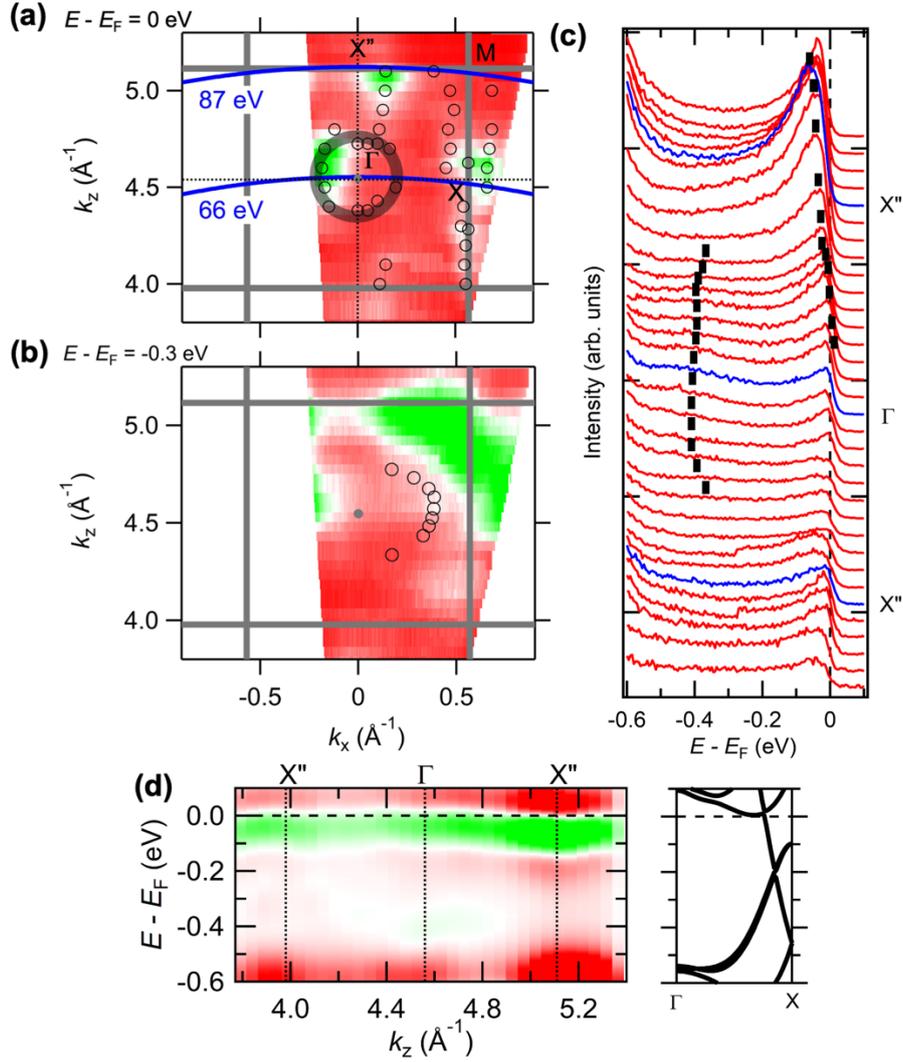

**Fig. S1** (a) ARPES intensity map showing $k_z$ vs $k_x$ at $E - E_F = 0$ eV, corresponding to Fermi surface, overlaid with the $k_F$ plots determined by fitting the MDC peaks (black circles). Blue lines show measured $k$ positions for the states near $E_F$ estimated for $h\nu = 66$ eV and for $h\nu = 87$ eV. The gray shaded circle is a guide for the eye. (b) ARPES intensity map showing $k_z$ vs $k_x$ at $E - E_F = -0.3$ eV, overlaid with the $k_F$ plots determined by fitting the MDC peaks. (c) EDCs along the X"ΓX" direction. Black solid lines show peak positions. (d) Second derivative of ARPES intensity map along the X"ΓX" direction. Right panel shows the band structure calculated within GGA.

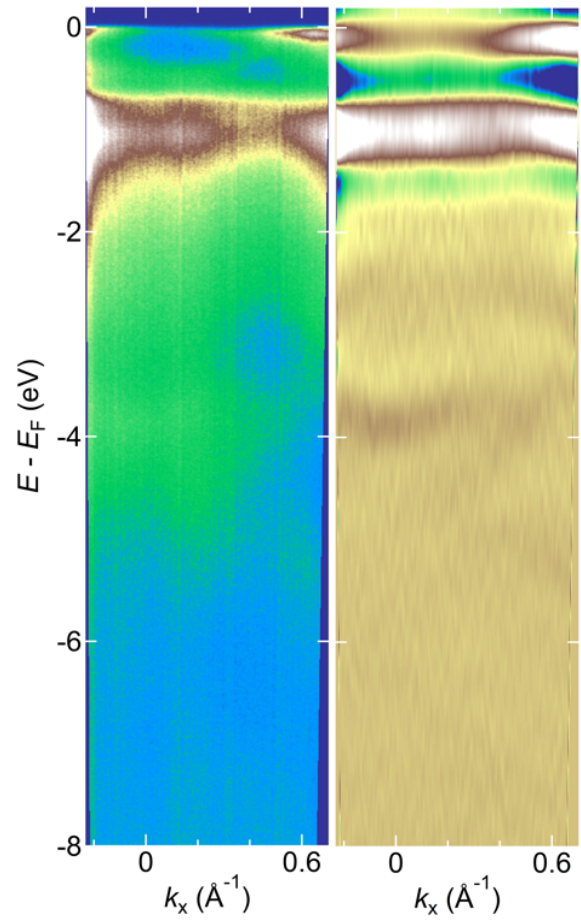

**Fig. S2** Valence band structure along the ΓX direction taken at $hv$ = 66 eV. (left) Raw ARPES intensity map and (right) second derivatives with respect to energy.

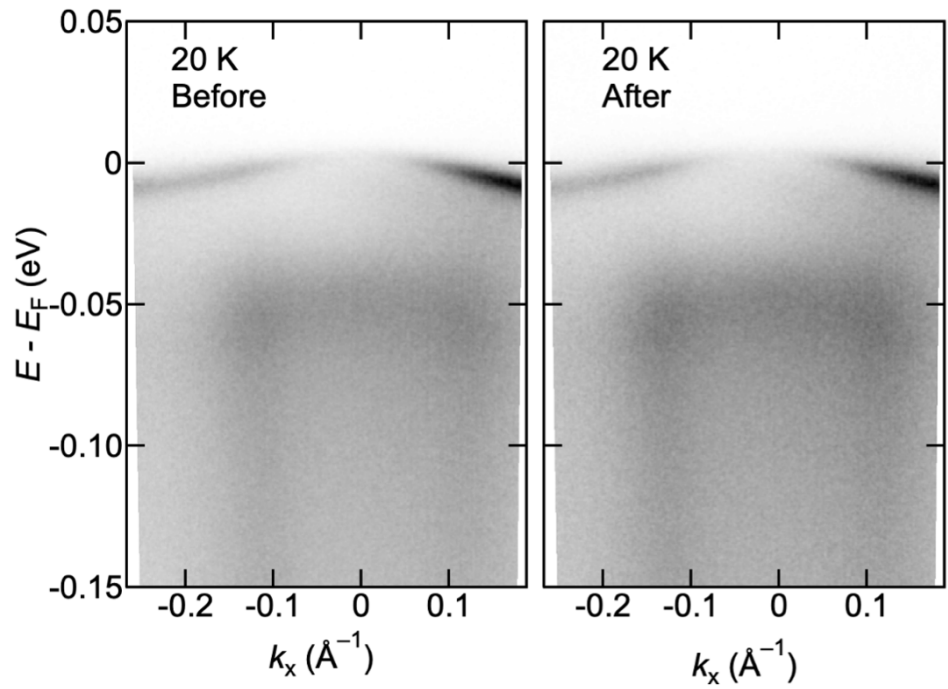

**Fig. S3** ARPES intensity maps taken with 7-eV laser before (left panel) and after (right panel) temperature-dependent measurements. The temperature-dependence measurements took approximately 14 hours.

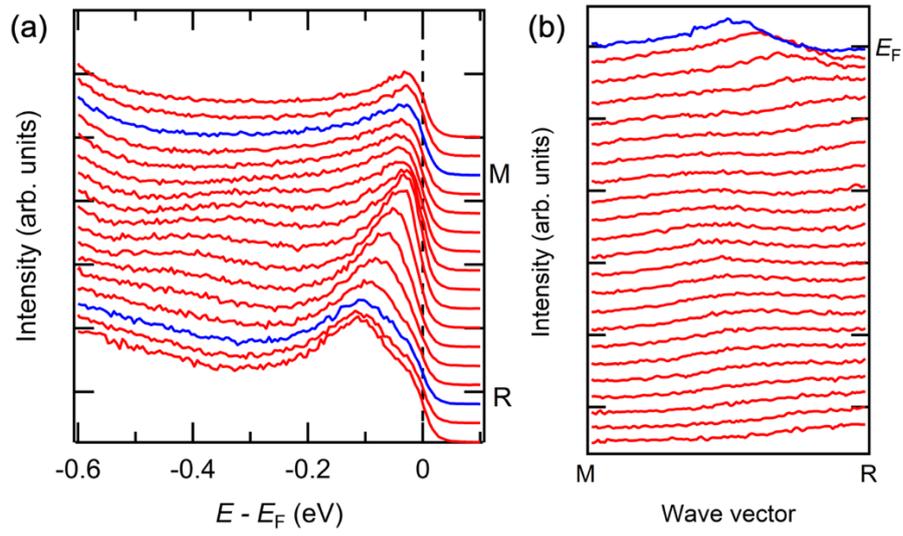

**Fig. S4** (a) EDCs along the MR direction. (b) MDCs along the MR direction shown in the range $E - E_F = -0.6$ eV to 0 eV,

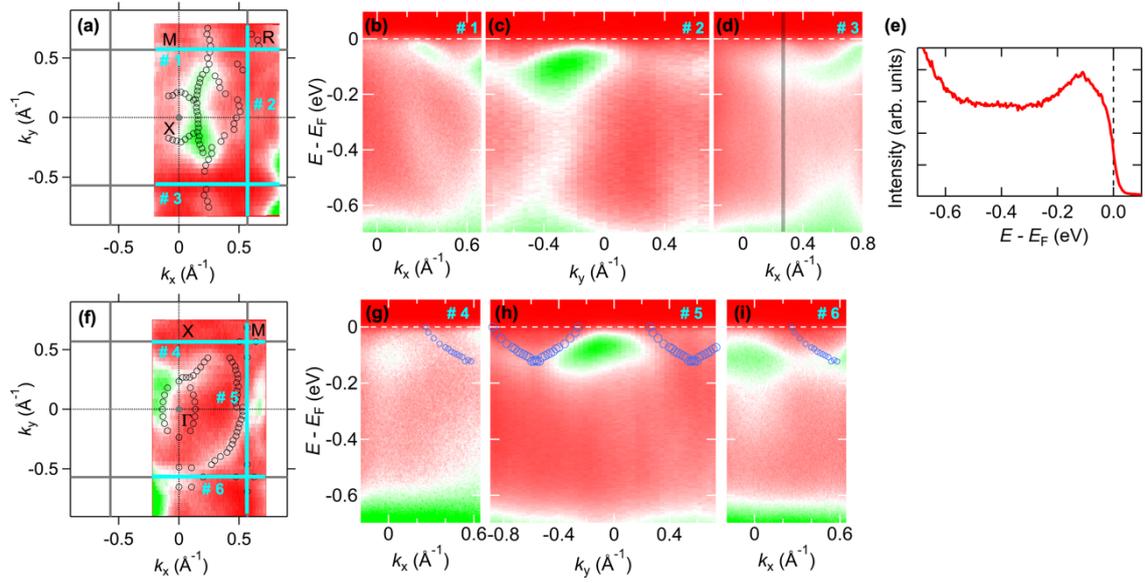

**Fig. S5** (a) Fermi surface in the XMRM plane. (b–d) *E-k* plot taken along #1–#3, respectively, as indicated by the light blue lines in (a). (e) Energy distribution curve taken along a gray line in (d). (f) Same as (a) but in the ΓXMX plane. (g–i) Same as (b–d) but along #4–#6, respectively, as indicated by the light blue lines in (f).

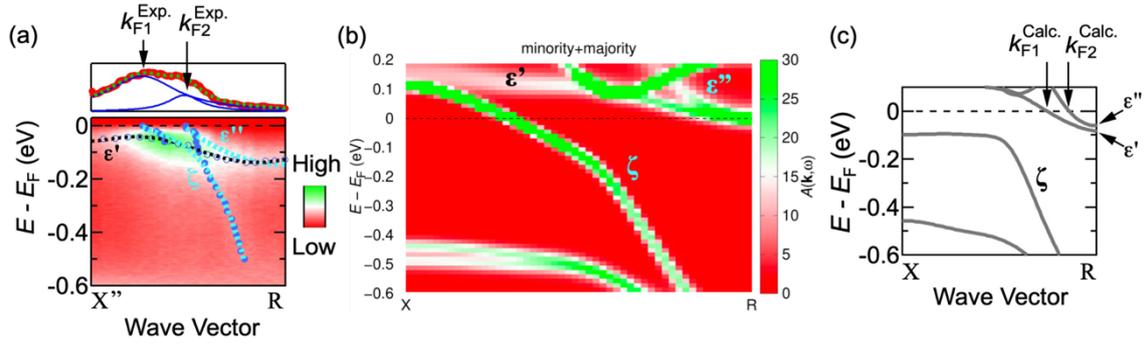

Fig. S6 (a) (top) Momentum distribution curve (MDC) at $E - E_F = 0$ with an energy window of ±0.01 eV (red dots). The green dotted line is the sum of the lorentzians (blue lines) corresponding to the modified theoretical bands crossing $E_F$. (bottom) ARPES intensity map measured along the X"R line overlaid with the band plots showing the peak positions of EDCs (white empty circles) and MDCs (white filled circles). Black and light blue dashed lines are guides for the eye, which is drawn based on the band picture demonstrated by DMFT calculation, as shown in (b). (b) Momentum-resolved spectral functions along the XR direction calculated within DFT+DMFT. Black dashed lines are guides for the eye. (c) Band structure along the XR line calculated within GGA with spin-orbit interactions.

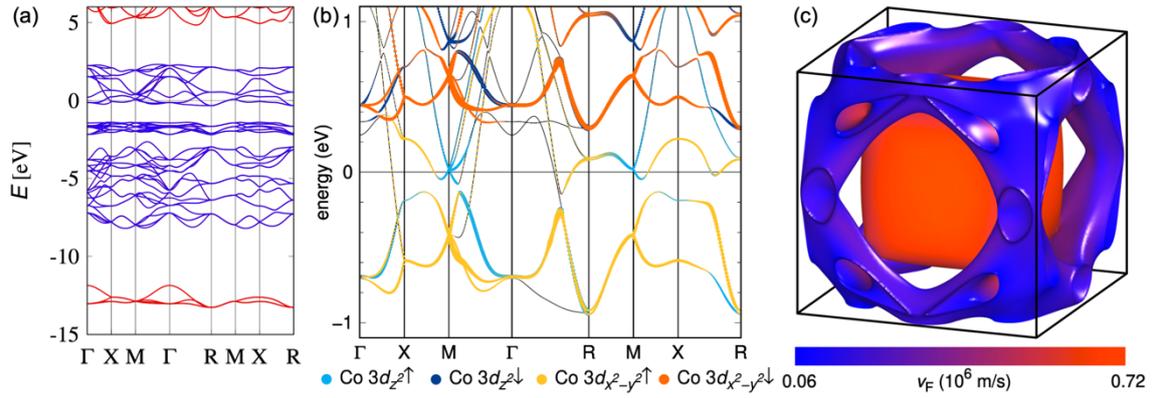

**Fig. S7** (a) Comparison of GGA band structure (red lines) and fit by the Wannier functions (blue points). The Fermi energy is set to zero. (b) Band structure showing majority and minority spin $e_g$ bands in the ferromagnetic state without exchange splitting reduced to 74%, and (c) corresponding Fermi surface.

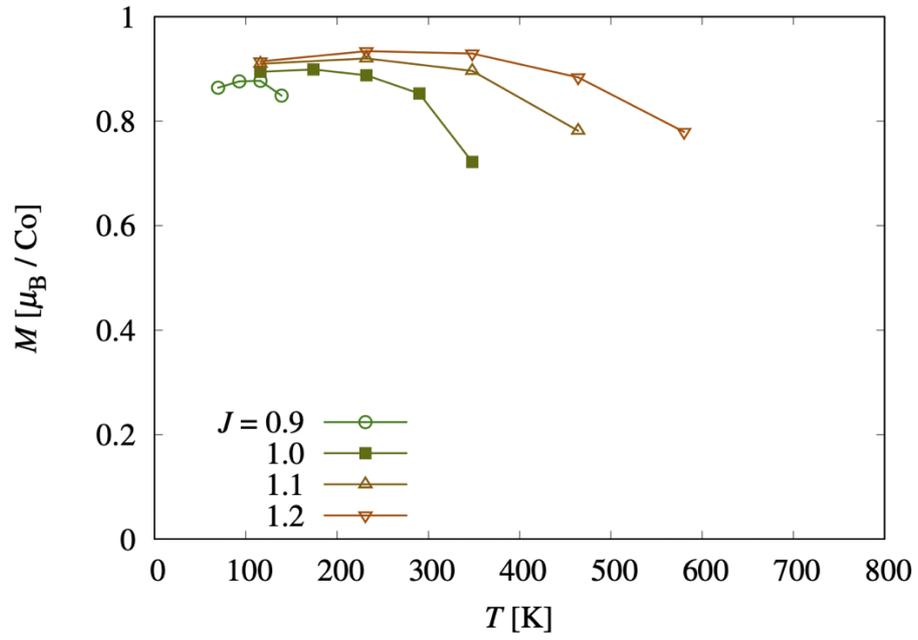

**Fig. S8** The ferromagnetic moment $M$ as a function of $T$ for several values of $J$ with $U =$ 4.

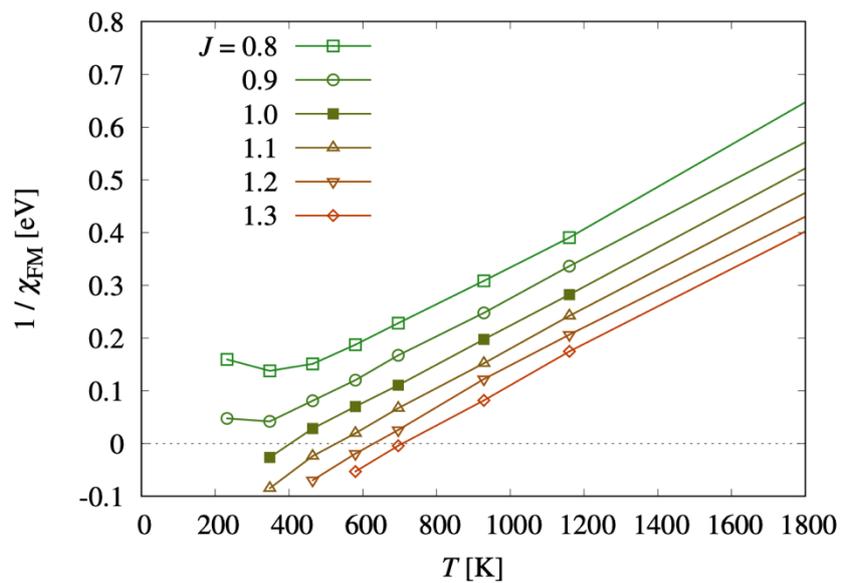

**Fig. S9** The temperature dependence of the inverse of the ferromagnetic susceptibility $\chi_{FM}$ for several values of $J$ for $U = 4$.

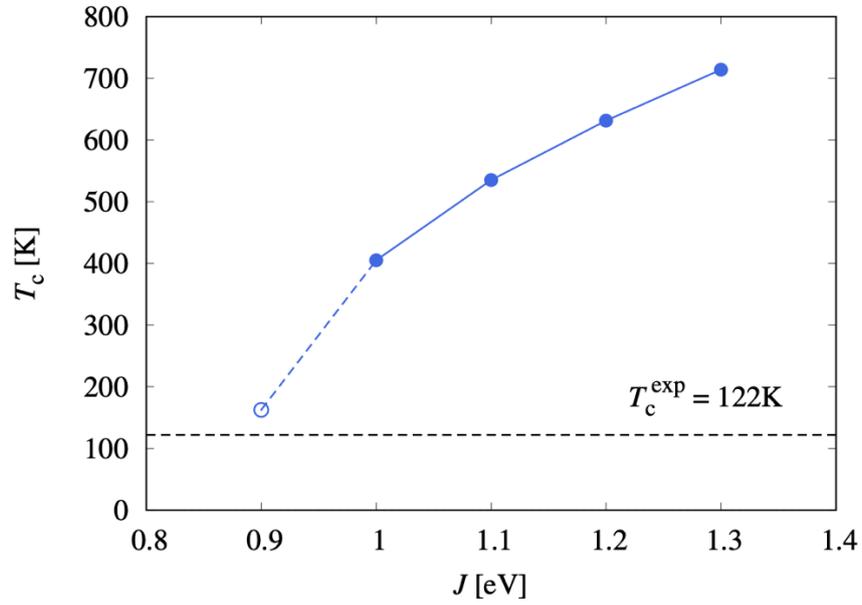

**Fig. S10** The ferromagnetic transition temperature $T_C$ as a function of $J$ for $U = 4$. The filled circles were obtained from the divergence of $\chi_{FM}$ and the open circle was obtained from vanishing $M$. The dashed horizontal line indicates the experimental value $T_C^{exp} = 122$ K.

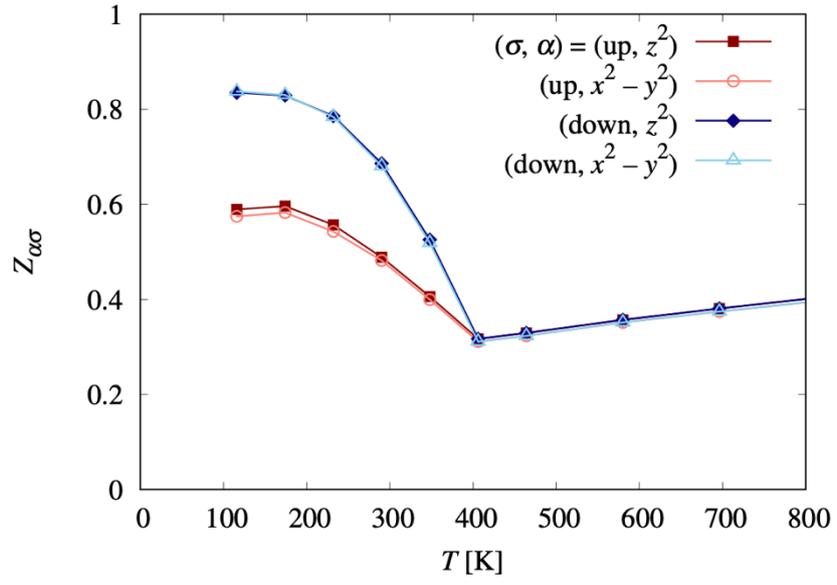

**Fig. S11** The spin- and orbital-dependent renormalization factor $Z_{\alpha\sigma}$ for $U = 4$ and $J = 1.0$.

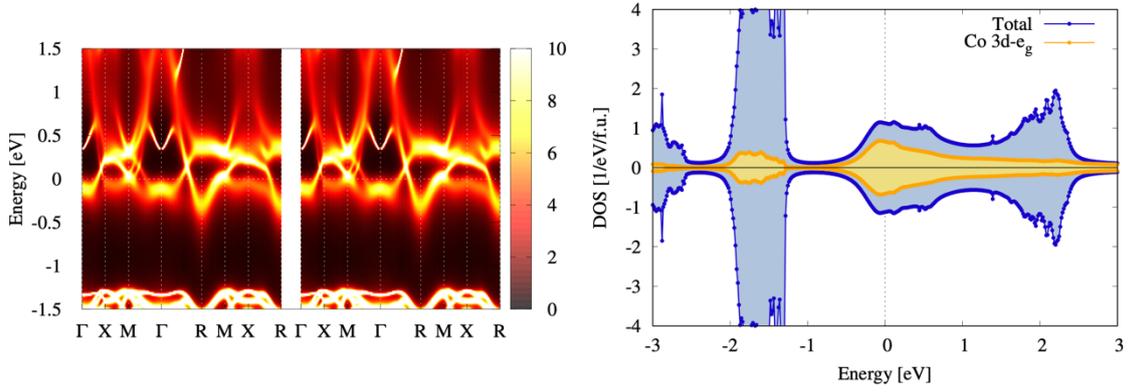

**Fig. S12** The single-particle excitation spectrum $A_\sigma(\boldsymbol{k},\omega)$ for $T/T_C = 1.16$ ($T = 464$ K), $U = 4$, and $J = 1.0$.

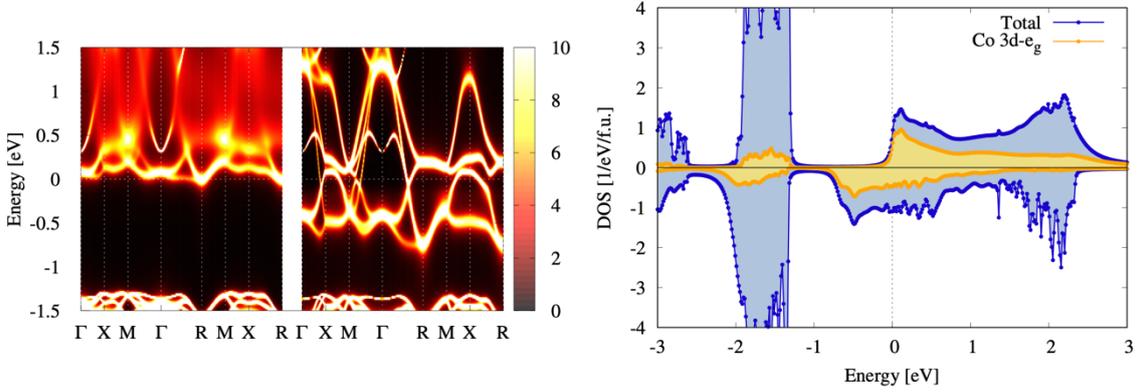

**Fig. S13** The single-particle excitation spectrum $A_\sigma(\boldsymbol{k},\omega)$ for $T/T_C = 0.86$ ($T = 348$ K), $U = 4$, and $J = 1.0$.

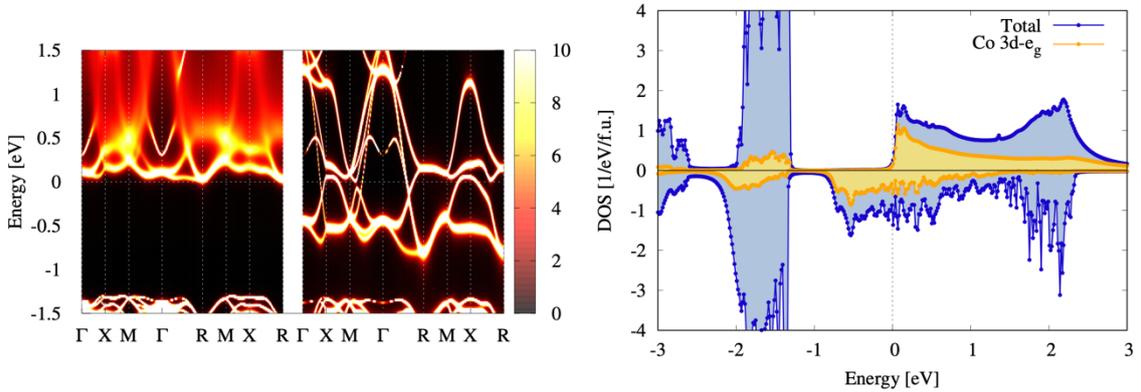

**Fig. S14** The single-particle excitation spectrum $A_\sigma(\boldsymbol{k},\omega)$ for $T/T_C = 0.72$ ($T = 290$ K), $U = 4$, and $J = 1.0$.

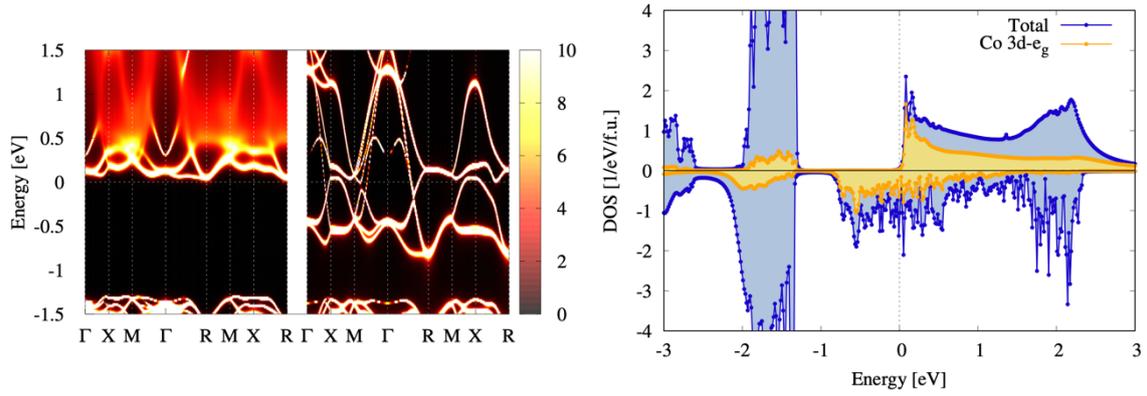

**Fig. S15** The single-particle excitation spectrum $A_\sigma(\boldsymbol{k},\omega)$ for $T/T_C = 0.58$ ($T = 232$ K), $U = 4$, and $J = 1.0$.

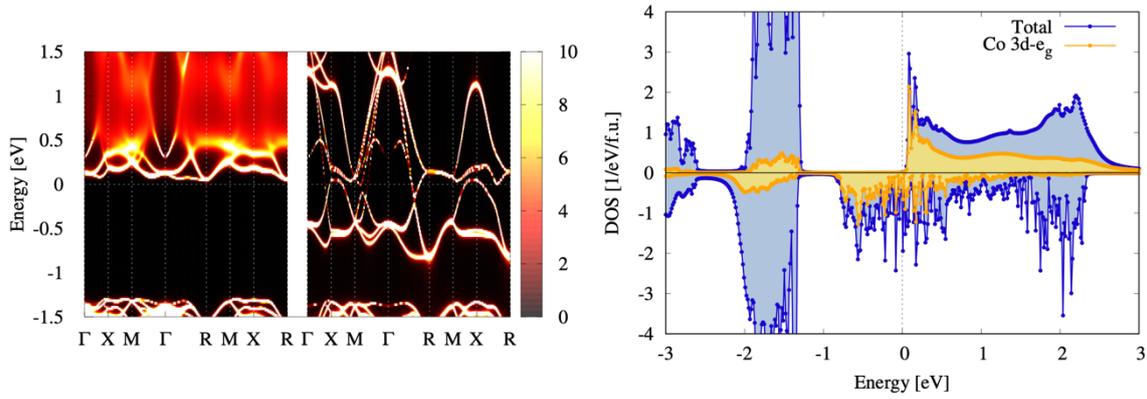

**Fig. S16** The single-particle excitation spectrum $A_\sigma(\boldsymbol{k},\omega)$ for $T/T_C = 0.43$ ($T = 174$ K), $U = 4$, and $J = 1.0$.

21